\pdfoutput=1
\newcommand*{\ATLASLATEXPATH}{latex/}
\documentclass[11pt, a4paper, british,texlive=2016,cernpreprint]{\ATLASLATEXPATH atlasdoc}

\usepackage[subfigure,backend=biber]{\ATLASLATEXPATH atlaspackage}
\usepackage{diagbox}
\usepackage{siunitx}
\usepackage{arydshln}

\usepackage{\ATLASLATEXPATH atlasbiblatex}

\usepackage{\ATLASLATEXPATH atlascontribute}

\usepackage{\ATLASLATEXPATH atlasphysics}
\usepackage{\ATLASLATEXPATH atlasbsm}
\usepackage{\ATLASLATEXPATH atlasprocess}

\graphicspath{{logos/}{figures/}}

\usepackage{Paper-defs}

\AtlasTitle{Search for long-lived charginos based on a disappearing-track signature in $\boldmath{pp}$ collisions at $\boldmath{\sqrt{s}=13}$~\TeV{} with the ATLAS detector}

\author{The ATLAS Collaboration}

\PreprintIdNumber{CERN-EP-2017-179}

\AtlasJournal{JHEP}
\AtlasJournalRef{JHEP 06 (2018) 022}
\AtlasDOI{10.1007/JHEP06(2018)022}

\AtlasAbstract{%
This paper presents a search for direct electroweak gaugino or gluino pair production with a chargino nearly mass-degenerate with a stable neutralino.
It is based on an integrated luminosity of 36.1~$\ifb${} of $pp$ collisions at $\sqrt{s} = 13$~\TeV{} collected by the ATLAS experiment at the LHC.
The final state of interest is a disappearing track accompanied by at least one jet with high transverse momentum from initial-state radiation or by four jets from the gluino decay chain.
The use of short track segments reconstructed from the innermost tracking layers significantly improves the sensitivity to short chargino lifetimes.
The results are found to be consistent with Standard Model predictions.
Exclusion limits are set at 95\% confidence level on the mass of charginos and gluinos for different chargino lifetimes. 
For a pure wino with a lifetime of about 0.2~ns, chargino masses up to 460~\GeV{} are excluded.
For the strong production channel, gluino masses up to 1.65~\TeV{} are excluded assuming a chargino mass of 460~\GeV{} and lifetime of 0.2~ns.
}

\hypersetup{pdftitle={ATLAS document},pdfauthor={The ATLAS Collaboration}}

\def\chinoone{\ensuremath{\mathchoice%
      {\displaystyle\raise.4ex\hbox{$\displaystyle\tilde\chi_1$}}%
         {\textstyle\raise.4ex\hbox{$\textstyle\tilde\chi_1$}}%
       {\scriptstyle\raise.3ex\hbox{$\scriptstyle\tilde\chi_1$}}%
 {\scriptscriptstyle\raise.3ex\hbox{$\scriptscriptstyle\tilde\chi_1$}}}}

\def\mdiff{\Delta m_{\chinoone}}

\begin{document}

\maketitle



\section{Introduction}\label{sec:introduction}

Supersymmetry (SUSY) \cite{Golfand:1971iw, Volkov:1973ix, Wess:1974tw, Wess:1974jb, Ferrara:1974pu, Salam:1974ig} is a space-time symmetry that relates fermions and bosons.
It predicts new particles that differ from their Standard Model (SM) partners by a half unit of spin.
If R-parity is conserved \cite{Farrar:1978xj}, SUSY particles are produced in pairs and decay such that their final products consist only of SM particles and the stable lightest supersymmetric particle (LSP).
In many supersymmetric models, the supersymmetric partners of the SM $W$ boson fields, the
wino fermions, are the lightest gaugino states. In this case, the lightest of the charged mass eigenstates, a chargino, and the lightest of the neutral mass eigenstates, a neutralino, are both almost pure wino and nearly mass-degenerate.
As a result, the lightest chargino can have a lifetime long enough that it can reach the ATLAS detector before decaying.
For example, anomaly-mediated supersymmetry breaking (AMSB) scenarios
\cite{Giudice:1998xp,Randall:1998uk} naturally predict a pure wino LSP, which is a
dark-matter candidate.
The mass-splitting between the charged and neutral wino ($\mdiff$) in such models is suppressed at tree level by the
approximate custodial symmetry; it has been calculated at the two-loop level to be around
160~\MeV~\cite{Ibe:2012sx}, corresponding to a chargino lifetime of about 0.2~ns~\cite{1126-6708-2003-03-045}.
This prediction for the value of the lifetime is actually a general feature of models with
a wino LSP: within the generated models of the ATLAS phenomenological Minimal
Supersymmetric Standard Model (pMSSM) scan~\cite{SUSY-2014-08} that have a wino-like LSP, about 70\% have a charged-wino lifetime between 0.15~ns and 0.25~ns. Most of the models in the other 30\% have a larger mass-splitting (and therefore the charged wino has a shorter lifetime) due to a non-decoupled higgsino mass.
The search presented here is sensitive to a wide range of lifetimes, from 10~ps to 10~ns, and reaches maximum sensitivity for lifetimes around 1~ns.

\begin{figure}[t]
  \centering
  \includegraphics[width=0.88\textwidth]{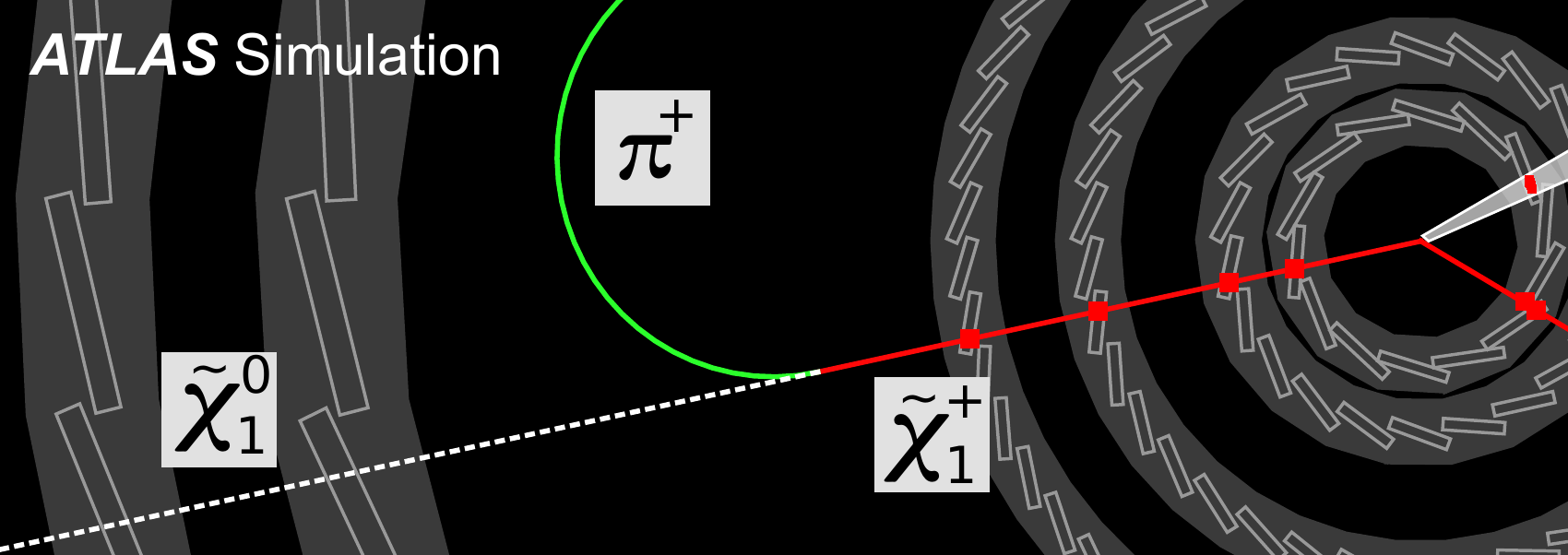}
  \caption{Illustration of a $pp\rightarrow \chinoonep \chinoonem + ~\mathrm{jet}$ event, with
      long-lived charginos. Particles produced in pile-up $pp$ interactions are not shown. The $\chinoonep$ decays into a low-momentum pion and a
          $\ninoone$ after leaving hits in the four pixel layers (indicated by red makers).}
  \label{figure:signal_simulation}
\end{figure}

The decay products of SUSY particles that are strongly mass-degenerate with the lightest neutralino leave little visible energy in the detector.
Thus, the corresponding searches represent a significant challenge for the LHC experiments.
If a charged SUSY particle produced in a high-energy collider had a relatively long lifetime, it would leave multiple hits\footnote{A hit is a space-time point which represents interactions between a particle and material in an active region of a particle detector.} in the traversed tracking layers before decaying, and could then be reconstructed as a track segment 
in the innermost part of the detector \cite{PhysRevD.55.330,Chen:1999yf,Asai2008185}. In the models considered in this paper, a
long-lived chargino decays into a pion and the LSP, a neutralino. The pion
emitted in the transition from the lightest chargino ($\chinoonepm$) to the lightest neutralino ($\ninoone$) typically has very low momentum
and it is not reconstructed in the detector. The neutralino is assumed to pass through the detector without interacting.
A track arising from a long-lived chargino can therefore {\em disappear}, i.e., leave
hits only in the innermost layers and no hits in the portions of the detector at higher
radii. Figure \ref{figure:signal_simulation} shows an example of a simulated signal event 
in which a long-lived chargino decays into a neutralino and a low-momentum pion in the ATLAS
detector. Searches for long-lived, massive, charged particles using measurements of ionization energy loss and timing information are also sensitive to long-lived charginos~\cite{SUSY-2016-03,SUSY-2016-02,CMS-EXO-15-010}, with a lower efficiency for selecting signals with lifetimes around 0.2~ns, relative to the disappearing-track signature. The disappearing-track signature provides the most sensitive search to date for SUSY models with charginos with $\mathcal{O}($ns$)$ lifetimes.

Previous searches for a disappearing-track signature were performed by the
ATLAS~\cite{SUSY-2013-01} and CMS~\cite{CMS-EXO-12-034} collaborations using the full dataset
of the LHC $pp$ run at a centre-of-mass energy of $\sqrt{s}=8$~\TeV{}. These searches
excluded chargino masses below 270~\GeV{} and 260~\GeV{} respectively, with a chargino proper lifetime ($\tau_{\chinoonepm}$) of 0.2~ns.
In the previous ATLAS analysis, a special tracking algorithm was used to reconstruct short tracks, and the search was sensitive to charginos decaying at radii larger than about 30~cm. A crucial improvement in the analysis described here is the use of even shorter tracks, called tracklets, which allows the reconstruction of charginos decaying at radii from about 12~cm to 30~cm. The use of these tracklets is possible thanks to the new innermost tracking layer \cite{ATLAS-TDR-19, IBL}
installed during the LHC long shutdown between Run 1 and Run 2. The use of shorter
tracklets significantly extends the sensitivity to smaller chargino lifetimes.

This paper is organised as follows. A brief overview of the ATLAS detector is given in
section~\ref{sec:detector}. In section~\ref{sec:overview}, the signal processes and
backgrounds are described and an overview of the analysis method is given. The data samples used in this analysis and the simulation model of the signal processes are described in section~\ref{sec:samples}.
The reconstruction algorithms and event selection are presented in section~\ref{sec:selection}. The analysis method is discussed in section~\ref{sec:analysis}. The systematic uncertainties are described in section~\ref{sec:uncertainties}. 
The results are presented in section~\ref{sec:results}. Section~\ref{sec:conclusion} is devoted to conclusions.

\section{ATLAS detector} \label{sec:detector}

ATLAS~\cite{PERF-2007-01} is a multipurpose detector with a forward-backward
symmetric cylindrical geometry, covering nearly the entire solid angle
around an interaction point of the LHC.\footnote{
ATLAS uses a right-handed coordinate system with its origin at the nominal interaction point in the centre of the detector. The positive $x$-axis is defined by the direction from the interaction point to the centre of the LHC ring, with the positive $y$-axis pointing upwards, while the beam direction defines the $z$-axis. Cylindrical coordinates $(r,\phi)$ are used in the transverse plane, $\phi$ being the azimuthal angle around the $z$-axis. The pseudorapidity $\eta$ is defined in terms of the polar angle $\theta$ by $\eta=-\ln\tan(\theta/2)$ and the rapidity is defined as $y = (1/2)\ln[(E+p_z)/(E-p_z)]$ where $E$ is the energy and $p_z$ the longitudinal
momentum of the object of interest.}
The inner tracking detector (ID) consists of pixel and micro-strip silicon detectors covering the pseudorapidity region of $|\eta|<2.5$, surrounded by a transition radiation tracker (TRT), which improves the momentum measurement and enhances electron identification
capabilities.
The pixel detector spans the radius range from 3~cm to 12~cm, the strip semiconductor tracker (SCT) from 30~cm to 52~cm, and the TRT from 56~cm to 108~cm.
The pixel detector has four barrel layers, and three disks in each of the forward and backward regions.
The barrel layers surround the beam pipe at radii of 33.3, 50.5, 88.5, and 122.5~mm, covering $|\eta|<1.9$.
These layers are equipped with pixels which have a width of 50~$\mu$m in the transverse direction.
The pixel sizes in the longitudinal direction are 250~$\mu$m for the first layer and 400 $\mu$m for the other layers. 
The innermost layer, the insertable B-layer \cite{ATLAS-TDR-19, IBL}, was added during the long shutdown between Run 1 and Run 2 and improves the reconstruction of tracklets by adding an additional measurement point close to the interaction point.
The ID is surrounded by a thin superconducting solenoid providing an axial 2~T magnetic
field and by a fine-granularity lead/liquid-argon (LAr) electromagnetic calorimeter
covering $|\eta|<3.2$. The calorimeters in the region of $3.1 < |\eta| < 4.9$ are
made of LAr active layers with either copper or tungsten as the absorber material. A
steel/scintillator-tile calorimeter provides coverage for hadronic showers in the central
pseudorapidity range of $|\eta|<1.7$. LAr hadronic end-cap calorimeters, which use lead as
absorber, cover the forward region of $1.5 < |\eta| < 3.2$. The muon spectrometer with an air-core toroid magnet system surrounds the calorimeters.
The ATLAS trigger system~\cite{TRIG-2016-01} consists of a hardware-based level-1 trigger followed by a
software-based high-level trigger.

\section{Analysis overview} \label{sec:overview}

\subsection{Signal processes} \label{sec:signal}

If the gluino mass is too large to yield a sizeable production cross-section,
electroweak-gaugino direct pair production could be the only gaugino
production mode within reach at LHC energies.
If the gluino mass is relatively light, however, gluino pair
production becomes the dominant process, and charginos can
be produced in cascade decays of the gluino.
For large mass separations between the gluino and the chargino, the
relatively large transverse momentum (\pT) transferred to the chargino
typically leads to higher kinematic selection efficiencies and larger
chargino decay radii relative to charginos from gaugino pair production.
Two complementary searches are described here: one targets direct
electroweak-gaugino pair production and the other targets gluino pair
production in which at least one long-lived chargino is produced in
the subsequent decay of the gluinos.
In both searches, events are
selected with a trigger based on the magnitude of the missing transverse
momentum in the event ($\met$).
A candidate event is required to have at least one ``pixel tracklet'', which is a tracklet with no associated SCT hits.
Candidate pixel tracklets are required to have $\pt>5$~\GeV.

\paragraph{Electroweak production:}\label{sec:ewkprod}

\begin{figure}[t]
  \centering
  \subfigure[$pp \rightarrow \chinoonepm \ninoone j$]{
    \includegraphics[height=0.35\textwidth]{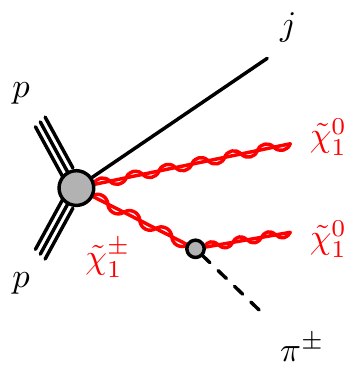}
  }
  \subfigure[$pp \rightarrow \chinoonepm \ninoone qqqq$]{
    \includegraphics[height=0.32\textwidth]{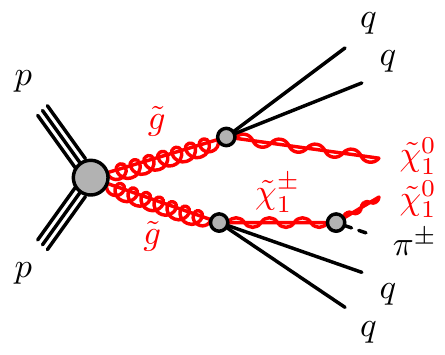}
  }
   \caption{Example diagrams of the benchmark signal processes used in this analysis.
      In the case of direct chargino/neutralino
      production (a), the signal signature consists of a long-lived
      chargino, missing transverse momentum and initial-state radiation.
         In the case of the strong channel (b),
             each gluino decays into two quarks and a chargino or neutralino. A long-lived
                chargino, missing transverse momentum and multiple quarks, which are observed as
                jets, are the signatures of this signal.}
   \label{figure:susy_diagram}
\end{figure}

This search targets the production processes $pp \rightarrow \chinoonepm
\ninoone j$ and $pp \rightarrow \chinoonep \chinoonem j$, where $j$ denotes an energetic
jet from initial-state radiation (ISR).
The presence of the ISR jet is required to ensure significant $\met$ and hence high trigger efficiency.
An example diagram for the $pp \rightarrow \chinoonepm \ninoone j$ process is presented in figure~\ref{figure:susy_diagram}(a).
The resulting signal topology is characterised by a high-\pt jet, large \met, and at least
one high-$\pt$ pixel tracklet.

\paragraph{Strong production:}\label{sec:strongprod}
This search targets gluino pair production with a long-lived chargino in the decay chains 
$pp \rightarrow \tilde{g}\tilde{g} \rightarrow qqqq \chinoonepm \ninoone$ 
and $pp \rightarrow \gluino\gluino \rightarrow qqqq\chinoonepm \chinoonepm$. These are typical decay modes in AMSB models.
An example diagram is shown in figure~\ref{figure:susy_diagram}(b). The signal topology
is characterised by four high-$\pt$ jets, large \met, and at least one high-$\pt$
pixel tracklet.

\subsection{Background sources} \label{sec:backgroundsource}

\begin{figure}[t]
  \centering
  \subfigure[]{
    \includegraphics[width=0.29\textwidth]{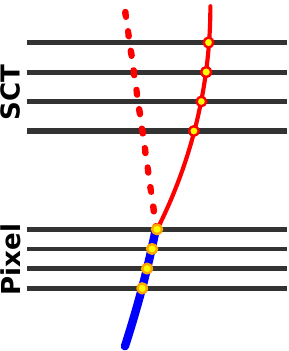}
    \label{fig:hadronSch}
  }
  \subfigure[]{
    \includegraphics[width=0.29\textwidth]{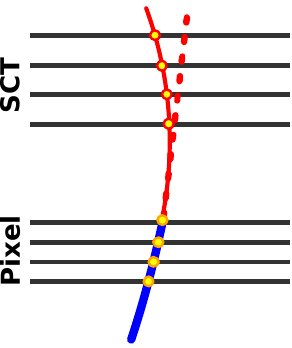}
    \label{fig:leptonSch}
  }
  \subfigure[]{
    \includegraphics[width=0.29\textwidth]{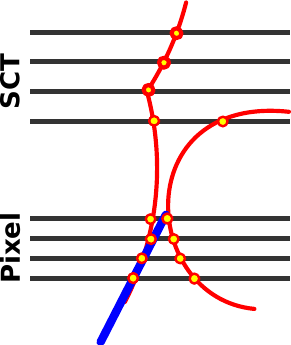}
    \label{fig:fakeSch}
  }
  \caption{Sketch of the different background components in the search with pixel
     tracklets. Thin solid and dotted red lines show trajectories of charged and neutral
        particles respectively. Thick blue lines show reconstructed pixel tracklets. (a) A
        hadron undergoing a hard scattering can yield track segments in the pixel and SCT
        detector that are not recognised as belonging to the same track, thus faking a
        pixel tracklet. (b) A lepton emitting hard photon radiation could be
        identified as a pixel tracklet through a similar mechanism. (c) A
        pixel tracklet can arise from a random combination of hits created by
        different particles in close proximity.}
  \label{figure:backgroundSchematic}
\end{figure}

The main SM background processes for the two analysis channels are top-quark pair production ($t\overline{t}$)
and $W$ boson production associated with hadron jets ($W$+jets) with subsequent decay $W\rightarrow e\nu, \tau\nu$. Hadrons
or leptons in these events can be reconstructed as
a pixel tracklet if they interact with the detector material and
any hits in the tracking detectors after the pixel detector are not
assigned to the reconstructed tracklet. Interactions that contribute to
this background include severe multiple-scattering, hadronic interactions or, in the case of leptons, bremsstrahlung, as shown in figure~\ref{fig:hadronSch} and \ref{fig:leptonSch}.
The other main category of background is from ``fake'' tracklets, which originate from random
combinations of hits from two or more particles, as shown in figure~\ref{fig:fakeSch}.

\subsection{Analysis method} \label{sec:analysis_in_short}

Candidate events are required to have large $\met$, at least one high-\pt jet, and at least one
isolated pixel tracklet.
A lepton-veto is used to suppress background events from $W$/$Z$ + jets and top-pair production processes.
Kinematic requirements, optimised for each channel, are applied to enhance the signal purity
in the event samples.
After selection, the search is performed by looking for an excess of candidate events in the
\pt distribution of pixel tracklets.
The shapes of the \pt spectrum for the background from hadrons,
muons, electrons, and fake tracklets are derived from
data using dedicated techniques for each background process. A fit to
the observed \pt distribution to extract the normalisation of the total background component and the signal strength is performed simultaneously in a low-$\met$ control region, two fake-tracklet control regions, and a high-$\met$
signal region. The regions are defined by the requirements described in section \ref{sec:selection} and in section \ref{sec:fit}.
The expected signal spectrum and yield are estimated from simulation and the measured
detector performance. Further details are given in section~\ref{sec:analysis}.

\section{Data and simulated event samples}\label{sec:samples}

The data used in this analysis were recorded by the ATLAS detector in 2015 and 2016. The $pp$
centre-of-mass energy was $13$~\TeV{} and the bunch spacing was 25~ns.
The mean number of $pp$ interactions per bunch crossing in the dataset was 14 in 2015 and 24 in 2016.

Events were selected by $\met$ triggers \cite{ATL-DAQ-PUB-2016-001}
with trigger thresholds varying from 70~\GeV{} to 110~\GeV{} depending on the data-taking period.
Data samples used to estimate the background contribution and to measure
tracking performance were selected using triggers requiring at least one
isolated electron ($\pt>24$--26~\GeV) or muon ($\pt>20$--26~\GeV).
After applying basic data-quality requirements, the data sample corresponds to an integrated luminosity of $36.1$~\ifb{}. 
The uncertainty in the combined 2015+2016 integrated luminosity is 3.2\%. It is derived, following a methodology similar to that detailed in ref.~\cite{DAPR-2013-01}, from a preliminary calibration of the luminosity scale using $x$--$y$ beam-separation scans performed in August 2015 and May 2016.

The simulated signal samples were generated assuming the minimal AMSB model \cite{Giudice:1998xp, Randall:1998uk} with $\tan\beta=5$,
    the sign of the higgsino mass term set to be positive, and the universal scalar mass set to $m_0=5$~\TeV{}.
The proper lifetime and
the mass of the chargino were scanned in the range from 10~ps to 10~ns and from 100~\GeV{}
to 700~\GeV{} respectively. For the strong production, samples were generated for gluino
masses ($m_{\tilde{g}}$) varying from 700~\GeV{} to 2200~\GeV{} with LSP mass from 200~\GeV{} to $m_{\tilde{g}}-100 $~\GeV{}.
The SUSY mass spectrum, the branching ratios and decay widths were calculated using {\scshape ISASUSY} 7.80~\cite{Paige:2003mg}.
The signal samples were generated with up to two extra partons in the matrix element using {\scshape MG5\_}a{\scshape MC@NLO}~2.3.3~\cite{Alwall:2014} at leading order (LO) interfaced to {\scshape Pythia 8.212}~\cite{Sjostrand:2014zea} for parton showering, hadronisation and SUSY particle decay. The NNPDF2.3LO~\cite{Ball:2014uwa} parton distribution function (PDF) set was used.
Renormalisation and factorisation scales were determined by the default dynamic scale choice of {\scshape MG5\_}a{\scshape MC@NLO}.
The {CKKW-L} merging scheme~\cite{Lonnblad:2011xx} was applied to combine tree-level matrix elements containing multiple partons with parton showers. The scale parameter for merging was set to a quarter of the mass of the wino for  wino-pair production or a quarter of the gluino mass for the strong production channel. 
The {\scshape A14}~\cite{ATL-PHYS-PUB-2014-021} set of tuned parameters with simultaneously optimised multiparton interaction and parton shower parameters was used for
the underlying event together with the NNPDF2.3LO PDF set.
Charginos were assumed to be stable in the event-generation step.

The cross-sections for the electroweak production are calculated at next-to-leading order (NLO) in the strong coupling constant using {\scshape Prospino2}~\cite{Beenakker:1996ch}.                              The cross-sections for the strong production are calculated in the same way as in the electroweak channel, adding the resummation of soft gluon emission at next-to-leading-logarithm accuracy (NLO+NLL)~\cite{Beenakker:2011fu}.
In both channels, an envelope of cross-section predictions is defined using the 68\% confidence level (CL) ranges of the {\scshape CTEQ6.6} PDF set~\cite{Nadolsky:2008zw}, including the $\alpha_\mathrm{S}$ uncertainty, and {\scshape MSTW2008} PDF set~\cite{Martin:2009iq}, together with variations of the factorisation and renormalisation scales by factors of two or one half.
The nominal cross-section value is taken to be the midpoint of the envelope and the uncertainty assigned is 
half of the full width of the envelope, following the PDF4LHC recommendations~\cite{Botje:2011sn}.
For the strong production mode, the branching ratio of the gluino decay is assumed to be 1/3 for each of the following decays: $\gluino \rightarrow qq\tilde{\chi}^0$, $\tilde{g} \rightarrow qq\tilde{\chi}^{-}$ and $\tilde{g} \rightarrow qq\tilde{\chi}^{+}$. Only first- and second-generation quarks ($d,u,s,c$) are considered. Direct electroweak-gaugino production is not considered in the strong channel.
The cross-section for the electroweak production, including at least one chargino, varies from 47~pb to 13~fb as the wino mass increases from 100~\GeV\ to 700~\GeV\,
with the uncertainty in the cross-section ranging from 8.6\% to 7.3\%. The cross-section
for gluino production varies from 3.5~pb to 0.36~fb as the gluino mass increases from
700~\GeV\ to 2200~\GeV\ with the uncertainty increasing from 14\% to 36\%.

The response of the detector to particles was modelled with the full ATLAS detector simulation \cite{SOFT-2010-01} based on {\scshape Geant4} \cite{Agostinelli:2002hh}.
Charginos were forced to decay into a pion and a neutralino in {\scshape Geant4}.
All simulated events were overlaid with
additional $pp$ interactions in the same and neighbouring bunch crossings (pile-up) simulated with the soft QCD processes of {\scshape Pythia 8.186} using the A2 set of tuned parameters~\cite{ATL-PHYS-PUB-2012-003} and the {\scshape MSTW2008LO} PDF set. The simulated events are reconstructed in the same way as the data, and are reweighted so that the distribution of the average number of collisions per bunch crossing matches the one observed in the data.

The $\met$ trigger efficiency is measured as a
function of the offline $\met$ using a data control sample consisting of events selected by the muon triggers and an additional offline selection designed to extract nearly pure $W \to \mu\nu$ events. For $\met>200$~\GeV, the trigger efficiency is almost 100\%. The measured trigger efficiency is used to directly estimate the probability for signal events to pass the trigger. 
The trigger efficiency for the direct electroweak production signal is about 20\%, depending on the assumed SUSY particle
masses. In the strong production search, the trigger efficiency is over 90\% when the mass difference between the gluino and the LSP is above 300~\GeV, and it decreases to about 55\% for a mass difference of 100~\GeV.

\section{Reconstruction and event selection}\label{sec:selection}

\subsection{Event reconstruction} \label{sec:object}

Primary vertices are reconstructed from two or more tracks with $\pt > 400$~\MeV.
When two or more vertices are reconstructed, the one with the largest sum of $\pt ^2$
of the associated tracks is used. Events are required to have at least one reconstructed primary vertex.

Jets are reconstructed from noise-suppressed energy clusters~\cite{PERF-2014-07} of calorimeter cells using an anti-$k_{t}$ algorithm~\cite{1126-6708-2008-04-063} with a radius parameter of 0.4 as implemented in the {\scshape FastJet} package \cite{Cacciari:2011ma}.
An area-based correction is applied to account for energy from additional $pp$ collisions based on an estimate of the pile-up activity in a given event \cite{Cacciari:2007fd}. 
Further corrections derived from the average jet response in simulation and data are used to calibrate the jet energies to the scale of their constituent particles~\cite{PERF-2016-04}.
Jets are required to have $\pt>20$~\GeV\ and $|\eta|<2.8$.
Additional selection criteria are applied to the tracks associated with jets~\cite{Aad:2015ina} with $\pt<60$~\GeV\ and $|\eta|<2.4$ to reduce the number of jets originating from pile-up interactions. 

Muon candidates are reconstructed by combining a track reconstructed by the muon spectrometer (MS track) with one recorded by the ID. They are required to
satisfy `Medium' quality requirements described in ref.~\cite{PERF-2015-10} and to have $\pt>10$~\GeV\ and $|\eta|<2.7$.

Electron candidates are reconstructed from energy clusters in the electromagnetic calorimeter with a matching track in the ID. They are required to satisfy
the `Loose' likelihood-based identification criteria described in ref.~\cite{ATLAS-CONF-2016-024}. 
They are further required to have transverse energy $E_{\mathrm{T}} > $10~\GeV\ and $|\eta|<2.47$.

After the requirements described above, ambiguities between candidate jets and leptons are
resolved as follows. First, any jet candidate which is within a distance $\Delta R\equiv\sqrt{(\Delta \eta)^2+(\Delta\phi)^2}=0.2$ of an electron candidate is discarded.
Second, if an electron (muon) candidate and a jet are found within $0.2 < \Delta R < 0.4$ (0.2 < $\Delta R <$min(0.4, 0.04 + 10~\GeV/$p_{\mathrm T}^{\mu}$)), the electron (muon) candidate is discarded and the jet is retained.
Finally, if a muon and a jet are found within $\Delta R < 0.2$, the muon is kept and the overlapping jet is ignored if fewer than three tracks with $\pT>500$~\MeV\ are associated with the jet.
The muon is ignored in the other case.
In addition, ambiguities between electrons and muons are resolved to avoid double counting: 
an electron is discarded if the electron candidate and a muon candidate share the same ID track. 

The offline missing transverse momentum~\cite{ATL-PHYS-PUB-2015-023} is calculated from the transverse momenta of selected
jets,
lepton candidates, and tracks compatible with the primary vertex but not associated with those leptons or jets.
Tracklets are not used in the $\met$ calculation.

Track reconstruction is performed in two stages. First, standard tracks, referred to as tracks in this paper, are reconstructed using a
standard algorithm \cite{ATL-PHYS-PUB-2009-002}. Tracks are required to have at
least seven hits in the silicon detectors \cite{ATL-PHYS-PUB-2015-018}.
A typical track for a high-$\pt$ charged particle which does not decay or scatter in the ID has four pixel hits, eight SCT hits, and 36 TRT hits at $\eta \approx 0$.
The track reconstruction is then rerun with looser criteria, requiring at least four
pixel-detector hits. The second reconstruction uses only pixel hits not associated with tracks as input, in order to find short tracks which are not reconstructed in the first step. Tracks reconstructed in the second step are referred to as tracklets.
The tracklets are then extrapolated to the SCT and TRT detectors, and any compatible hits
are assigned to the tracklet candidate.
Tracklets are required to have $\pt>5$~\GeV, $|\eta| < 2.2$, and their longitudinal impact
parameter\footnote{The transverse impact parameter is defined as the distance of closest approach in the transverse plane between a track and the centre of the luminous region. A correction is applied to take into account the tilt of the luminous region with respect to the $z$-axis. The longitudinal impact parameter corresponds to the $z$-coordinate distance between the point along the track at which the transverse impact parameter is defined and the primary vertex.} $|z_0|$, must be smaller than 10~mm.
Figure~\ref{figure:tracking_efficiency} shows the reconstruction efficiency for simulated charginos as a
function of the chargino decay radius, where requirements described later in this
section are not applied to compute the efficiency except for the
disappearance condition for pixel tracklets.
By using pixel tracklets rather than tracks, the reconstruction efficiency is improved significantly for
charginos decaying at radii less than 300~mm.
For charginos with a lifetime of 0.2~ns, which have a mean decay radius of 6~cm, the probability to reconstruct a pixel tracklet is 5--10\%; this tracklet reconstruction efficiency is a factor of ten greater than the efficiency obtained using tracks. The inefficiency in reconstructing pixel tracklets for charginos with a lifetime of 0.2~ns is largely due to charginos which decay before reaching the fourth layer of the pixel detector.

\begin{figure}[tbhp]
  \centering
     \includegraphics[width=0.8\textwidth]{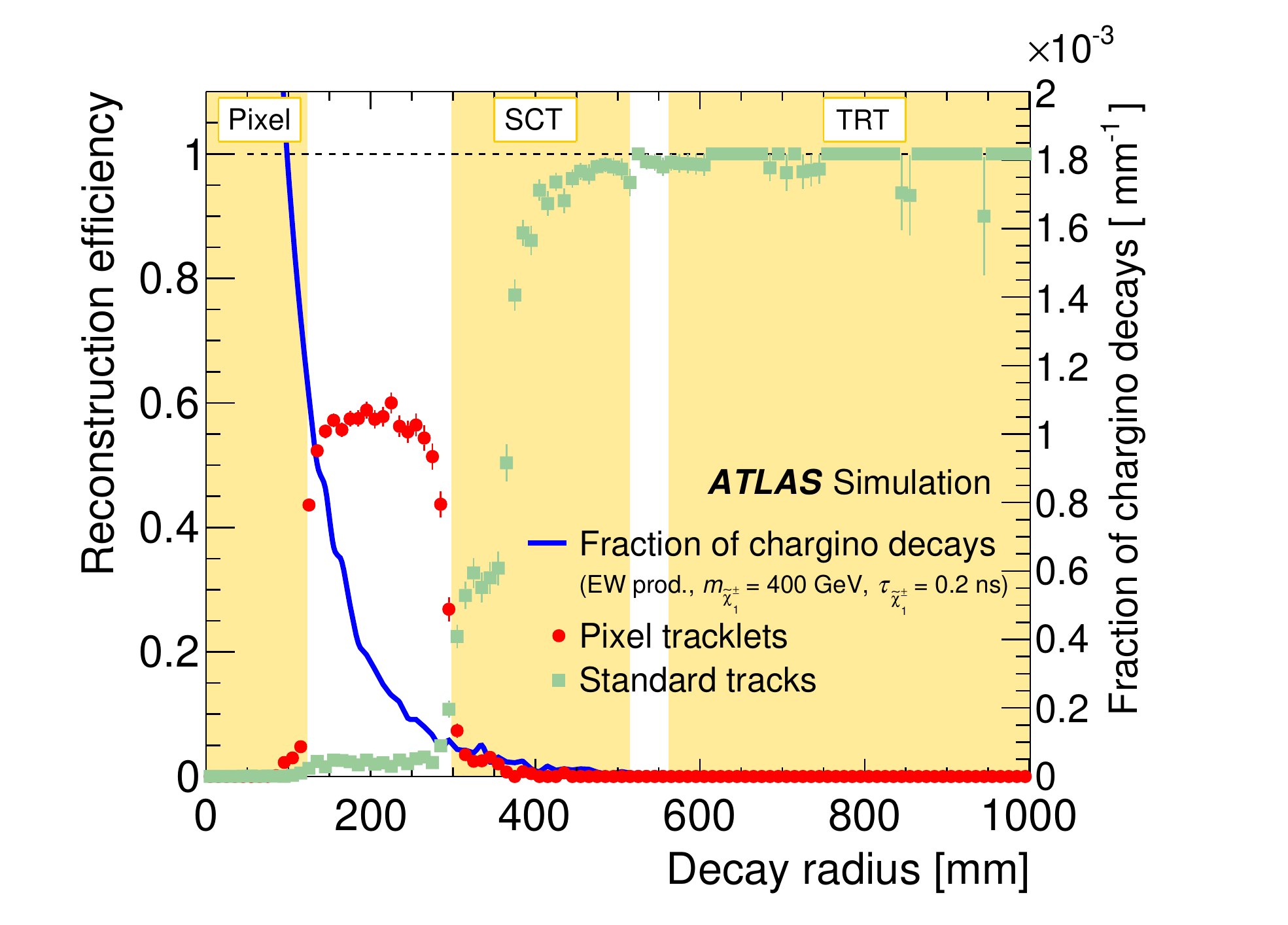}
  \caption{Chargino reconstruction efficiency as a function of decay radius.
    The reconstruction efficiency of pixel tracklets before applying the fake-rejection criteria is shown in red, while that obtained with the standard tracking algorithm is shown in green. 
    The error bars show statistical uncertainties in the estimation.
    Also shown in blue, on the right axis, is the distribution of the decay radius for charginos with a lifetime of 0.2~ns. 
    The yellow shaded regions correspond to the coverage of each detector.}
  \label{figure:tracking_efficiency}
\end{figure}

To reduce contributions from tracklets from background processes, the following requirements are applied to the tracklets:

\begin{itemize}
  \item[(1)] {\bfseries Isolation and \pt requirements:} 
  The separation $\Delta R$ between the
  tracklet and any jet with $\pt > 50$~\GeV{} or any reconstructed MS track
must be greater than 0.4. The candidate tracklet is required to be isolated.
  A track or tracklet is defined as isolated when
  the sum of the transverse momenta of all standard ID tracks with $\pt > 1$~\GeV\ and $|z_0 \mathrm{sin}(\theta)|<$ 3.0~mm in a cone of $\Delta R = 0.4$ around the
  track or tracklet, not including the $\pt$ of the candidate track or tracklet, divided by the track or tracklet \pT, is smaller than 0.04.
The candidate tracklet must have $\pt > 20$~\GeV, and the $\pt$ must be the highest
among isolated tracks and tracklets in the event.

   \item[(2)] {\bfseries Geometrical acceptance:} The tracklet must satisfy $0.1 < |\eta| < 1.9$.

   \item[(3)] {\bfseries Quality requirement:}  The tracklet is required to have hits on
  all four pixel layers.
  The number of pixel holes, defined as missing hits in modules in which at least one is expected
  given the detector geometry and conditions, must be
  zero.  The number of low-quality hits\footnote{A hit is categorised as low quality
         when the single-hit position uncertainty is large, or the hit position is far
            from the reconstructed tracklet.
} associated with the tracklet must be zero. 
Furthermore, tracklets must satisfy requirements on the significance of the transverse impact parameter, $d_0$
, $|d_0|/\sigma(d_0) <$ 2 (where $\sigma(d_0)$ is the uncertainty in the $d_0$ measurement), and
 $|z_0 \mathrm{sin}(\theta)|<$ 0.5~mm.                        
            The $\chi^{2}$-probability of the fit is required to be larger than 10\%.

  \item[(4)] {\bfseries Disappearance condition:} The number of SCT hits associated with the tracklet must be zero.
\end{itemize}

The isolation and quality requirements are mainly useful in rejecting fake tracklets, which have a flat distribution in impact parameter.
The requirement on $\eta$ excludes tracklets with $\eta \sim 0$, where the muon
spectrometer has low efficiency. Including tracklets in this region would
increase the background significantly, as the lepton rejection is less efficient.
Tracklets with $|\eta| >$ 1.9 are rejected because the probability of a particle scattered by detector material to be reconstructed as a
tracklet increases at high $|\eta|$. The disappearance condition is used to identify tracklets which arise from particles decaying between the pixel and the SCT detectors.

\subsection{Event selection} \label{sec:event_selection}
Events are selected by applying requirements on the event kinematics.
The selection requirements for the signal regions for the two different production channels are described below.

\paragraph{Event preselection:}
Common selection criteria are applied in the two searches.
At least one pixel tracklet must satisfy all the requirements described in section \ref{sec:object}.
To ensure good data quality, an event is rejected when the jet with the highest $\pt$ in the event passes the `BadTight' \cite{ATLAS-CONF-2015-029} selection or at least one jet passes the `BadLoose' \cite{ATLAS-CONF-2015-029} selection, which is used to reduce jets originating from detector noise and non-collision background.
Events containing a muon, before ambiguity removal between muons and jets, with momentum uncertainty $\sigma(q/p)/|q/p|>0.2$ are also rejected, where $q$ and $p$ are the electric charge and the magnitude of the momentum of the muon.
To suppress contributions from top-quark-pair ($t\overline{t}$) and $W/Z+\mathrm{jets}$ production processes, candidate events are required to have no electron and no muon candidates (lepton veto). 

\paragraph{Electroweak chargino production:}

Events are required to have at least one jet with
$\pt>140$~\GeV\ and $\met>140$~\GeV\ ($90~\GeV < \met < 140~\GeV$) in the high- (low-) \met region to discriminate the signal from SM processes.
In order to further suppress the multijet background, the difference in azimuthal angle
($\Delta \phi$) between the missing transverse momentum and each of the up to four highest-\pT jets with $\pT > 50$~\GeV\ is required to be larger than 1.0.

\paragraph{Strong production:}

Candidate events are required to have a jet with $\pt>100$~\GeV,
at least two additional jets with $\pt>50$~\GeV\ and $\met>150$~\GeV\ ($100~\GeV < \met < 150~\GeV$) in the high- (low-) \met region to discriminate the signal from SM processes.
In order to further suppress the multijet background, the $\Delta \phi$ between the missing transverse momentum
and each of the up to four highest-\pT jets with $\pt>50$~\GeV\ is required to be larger than 0.4.

\subsection{Signal acceptance and efficiency} \label{sec:eff}

The number of events observed in data and the expected number of signal events for two representative signal points are shown in table~\ref{table:cutflow}, for the selection described above. No generator-level requirements are applied to signal events. Therefore, events are counted in which the chargino decays before reaching the fourth pixel layer but an isolated track or a tracklet from an SM particle or from a random combination of hits is reconstructed. Such events are rejected by the isolation requirement, the geometrical acceptance or the quality selection.

\begin{table*}[t]
  \centering
  \begin{tabular}{l|rS[round-mode=off,group-minimum-digits=5]@{\hskip -0.45cm}l|rS[round-mode=off,group-minimum-digits=5]@{\hskip -0.19cm}l}
    \hline
    \hline
 Selection requirement  & \multicolumn{3}{c|}{Electroweak channel} & \multicolumn{3}{c}{Strong channel}  \\
                    & Observed & \multicolumn{2}{c|}{Expected signal} & Observed & \multicolumn{2}{c}{Expected signal}\\
    \hline
    Trigger                         & \num[group-minimum-digits = 5]{434559704} & 1276  & (0.20)   & \num[group-minimum-digits = 5]{434559704} & 285   & (0.98) \\
    Jet cleaning                    & \num[group-minimum-digits = 5]{288498579} & 1181  & (0.19)   & \num[group-minimum-digits = 5]{288498579} & 282   & (0.97) \\
    Lepton veto                     & \num[group-minimum-digits = 5]{275243946} & 1178  & (0.19)   & \num[group-minimum-digits = 5]{275243946} & 278   & (0.95) \\
    \met and jet requirements       & \num[group-minimum-digits = 5]{2697917  } & 579.1 & (0.092)  & \num[group-minimum-digits = 5]{537861   } & 202   & (0.69) \\
    \hdashline                                                                                   
    Isolation and $\pt$ requirement & \num[group-minimum-digits = 5]{464524   } & 104.2 & (0.017)  & \num[group-minimum-digits = 5]{107381   } & 43.6  & (0.15) \\
    Geometrical $|\eta|$ acceptance & \num[group-minimum-digits = 5]{339602   } & 83.6  & (0.013)  & \num[group-minimum-digits = 5]{77675    } & 36.4  & (0.13) \\
    Quality requirement             & \num[group-minimum-digits = 5]{6134     } & 29.6  & (0.0047) & \num[group-minimum-digits = 5]{1337     } & 13.9  & (0.048) \\
    Disappearance condition         & \num[group-minimum-digits = 5]{154      } & 24.1  & (0.0038) & \num[group-minimum-digits = 5]{35       } & 11.0  & (0.038) \\
    \hline
    \hline
  \end{tabular}
  \caption{Summary of the selection criteria, and the corresponding observed number of events in data as well as the expected
     number of signal events in simulation for two benchmark models: a chargino produced in direct electroweak production with {\small
    ($m_{\chinoonepm}$, $\tau_{\chinoonepm}$) = (400~\gev{}, 0.2~ns)} and a chargino produced in the strong channel with {\small ($m_{\gluino}$, $m_{\chinoonepm}$, $\tau_{\chinoonepm}$)  = (1600~\gev{}, 500~\gev, 0.2~ns)}
 in the high-$\met$ region. The expected number of signal events is normalised to 36.1~$\ifb$.
    The signal selection efficiencies are also shown in parentheses.
    The first row shows the number of events after the application of detector and data quality conditions. Requirements below the dashed line are applied to tracks and tracklets.}
  \label{table:cutflow}
\end{table*}

To facilitate reinterpretation, the signal efficiency and generator-level acceptance are shown in table~\ref{table:eff_acc} for a few signal models with the following definitions.
A generator-level event kinematic volume for electroweak production
is defined as: 1) $\met>140$~\GeV, 2) at least one jet with $\pt>140$~\GeV, 3) $\Delta
\phi > 1.0$ between the missing transverse momentum and each of the up to four highest-\pT jets with $\pT > 50$~\GeV, and 4) no electrons or muons. For strong production, the event requirements are: 1) $\met>150$~\GeV, 2) at least one jet with $\pt>100$~\GeV\ and
at least two more jets with $\pt>50$~\GeV, 3) $\Delta \phi > 0.4$ between the missing transverse
momentum and each of the up to four highest-\pT jets with $\pT > 50$~\GeV, and 4) no electrons or muons.
The generator-level missing transverse momentum is defined as the vector sum of the \pT of neutrinos, neutralinos and charginos, as the \pt of a tracklet is not used in the reconstruction of missing transverse momentum.
The generator-level jets are built using the anti-$k_{t}$ algorithm with a radius parameter of 0.4, taking as input all particles, except for muons, neutrinos, neutralinos and charginos, with $c\tau>10$~mm.
The fraction of chargino events passing this generator-level kinematic selection is shown for several signal points as ``event acceptance''. The ``event efficiency'' is defined as the ratio of the number of reconstructed events which pass the requirements defined in section~\ref{sec:event_selection} (including the trigger requirement) to the number of events which fall into the generator-level acceptance volume defined above. The event efficiency does not include any requirement on tracklets.
The event efficiency can be greater than unity because an event which is not in the generator-level kinematic volume can pass the selection after reconstruction due to reconstruction resolutions.

The full selection efficiency must also consider the probability of reconstructing in the event at
least one tracklet that satisfies the four tracklet selection criteria defined
in section~\ref{sec:object}, and has a reconstructed \pT above 100~\GeV. This is quantified in table~\ref{table:eff_acc} based on a generator-level tracklet selection. To be accepted as a tracklet at generator level, a chargino must meet the following criteria: 1) $\pt > 20$~\GeV, 2) $0.1 < |\eta| < 1.9$, 3) 122.5 mm $<$ decay position $<$ 295 mm, where the decay position is the cylindrical radius relative to the origin, and 4) $\Delta R > 0.4$ between the chargino and each of the up to four highest-\pT jets with $\pT > 50$~\GeV. The fraction of produced charginos which pass this generator-level selection, in events which pass the event-level selection requirements, is shown as ``tracklet acceptance.'' Given that a tracklet passes these requirements at generator level, the probability for it to pass the full pixel tracklet selection at reconstruction level is defined as ``tracklet efficiency,'' and the probability for such a reconstructed tracklet
to have $\pt > 100$~\GeV\ is shown independently. The selection efficiency is shown per tracklet, and therefore for events with two charginos, the full probability of selecting the event must take into account the probability of at least one of the tracklets passing both acceptance and efficiency.\footnote{If $E_A$ is the event acceptance, $E_E$ is the event efficiency, $T_A$ is the tracklet acceptance, $T_E$ is the tracklet efficiency, and $T_p$ is the tracklet $\pt$ efficiency, then for an event with $N$ charginos, the probability of having at least one reconstructed, selected tracklet with $\pt > 100$ GeV in an event can be calculated as: $E_E \times E_A \times (1-(1-T_A \times T_E \times T_P)^{N})$.} For models in which the signal $\pt$ spectrum differs significantly from that of the charginos considered here, the momentum resolution of the tracklets, as shown in figure~\ref{figure:smearingFunction_fitting}, must be taken into account to correctly estimate the probability of reconstructing $\pt > 100$~\GeV\ (see section ~\ref{sec:background}). The tracklet efficiency depends on the soft jet activity around a chargino, and therefore differs between the charginos produced via electroweak and strong production mechanisms.

\begin{table}[t]
  \centering
  \begin{tabular}{c|c|c|c|c|c| c}
  \hline
    \hline
\multicolumn{2}{c|} {Signal model} & \multicolumn{2}{c|} {Event}  & \multicolumn{3}{c} {Tracklet} \\
    \hline 
Mass [\GeV] & Lifetime [ns] & Acceptance & Efficiency & Acceptance & Efficiency & $P$\\
  \hline
   \multicolumn{7}{c}{Electroweak production}\\
  \hline
   $m_{\tilde{\chi}^{\pm}_1}$=400                            & 0.2&\num[round-precision=2]{0.0934731}&\num[round-precision=2]{1.03090} & \num[round-precision=2]{0.0666299}& \num[round-precision=2]{0.470852}& \num[round-precision=2]{0.571504}\\
   $m_{\tilde{\chi}^{\pm}_1}$=600                            & 0.2&\num[round-precision=2]{0.115797} &\num[round-precision=2]{1.04731} & \num[round-precision=2]{0.0504342}& \num[round-precision=2]{0.483321}& \num[round-precision=2]{0.569029}\\
   $m_{\tilde{\chi}^{\pm}_1}$=600                            & 1.0&\num[round-precision=2]{0.114653} &\num[round-precision=2]{1.03177} & \num[round-precision=2]{0.19676}  & \num[round-precision=2]{0.470673}& \num[round-precision=2]{0.565923} \\ 
    \hline
   \multicolumn{7}{c}{Strong production}\\
  \hline
   {\small $m_{\tilde{g}}$=1600, $m_{\tilde{\chi}^{\pm}_1}$=500}& 0.2&\num[round-precision=2]{0.71} & 0.97 & \num[round-precision=2]{0.104}& 0.38 & 0.55 \\
   {\small $m_{\tilde{g}}$=1000, $m_{\tilde{\chi}^{\pm}_1}$=900}& 0.2&\num[round-precision=2]{0.18} & 0.93 & \num[round-precision=2]{0.0281} & 0.36 & 0.55 \\
  \hline
    \hline
  \end{tabular}
  \caption{The event and tracklet generator-level acceptance and selection efficiency for
      a few signal models studied in this search. The last column shows
the probability ($P$) for a reconstructed tracklet to have $\pt$ greater than 100~\GeV. For details, see text. }
  \label{table:eff_acc}
\end{table}

\section{Signal and background estimation} \label{sec:analysis}

An unbinned likelihood fit is performed on the $\pt$ distribution of the pixel tracklets in a wide \pt range, \pt $> 20~\GeV$.
Most of the signal events are expected to be in the high-\met region. The contamination of
signal in the low-$\met$ region is at the level of 3\%, and this region is used to constrain the fake-tracklet $\pt$ spectrum.

\subsection{Background templates} \label{sec:background}

Templates for background components are estimated from data.
The \pt spectra of hadrons and leptons scattered by the ID material are estimated from the
\pt distribution of tracks associated with non-scattered hadrons and leptons, selected in
dedicated control samples (as detailed in section~\ref{sec:hadronbg} and in section~\ref{sec:lepton}), by smearing them to
take into account the poor \pt resolution of pixel tracklets.
The \pt spectrum shape of the fake-tracklet component is also obtained in a dedicated control region (as detailed in section~\ref{sec:fakes}).

\subsubsection{Smearing function} \label{sec:smearing}
A smearing function to translate from track to tracklet momentum resolution is extracted from $Z \rightarrow \mu\mu$ events in data by
re-fitting the muon candidate track using only the hits in the pixel detector.
The $Z \rightarrow \mu\mu$ events are selected by single-muon triggers
and by requiring two opposite-charge muons with a difference in azimuthal
angle larger than 1.5, and with an invariant mass between 81~\GeV\ and 101~\GeV.
The $q/\pt$ resolution of pixel tracklets is calculated from
the distribution of the difference between the $q/\pt$ of the pixel tracklet and the original
track. This distribution is shown in figure~\ref{figure:smearingFunction_fitting}(a). 
The $q/\pt$ difference distribution is modelled by the following empirical formula:
\begin{eqnarray}
f(z) &=&
\begin{cases}
\exp( \alpha (z + \alpha/2) ) & (z < -\alpha) \\
\exp( - z^2/2 )                 & (-\alpha < z < \alpha), \\
\exp(-\alpha (z - \alpha/2) ) & (z > \alpha) \\
\end{cases} \nonumber \\
z &=& \frac{\Delta (q/\pt) - \beta}{\sigma},
\label{eq:CrystalballLikeFunction}
\end{eqnarray}
where $\alpha$, $\beta$ and $\sigma$ are parameters representing the slope of the
tail part, and the mean and resolution of the core part of the distribution respectively.
The measured $q/\pt$ resolution of pixel tracklets, for \pt much larger than 10~\GeV{}, is 13.2~\tev${}^{-1}$,
    which is ten times larger than that of tracks with more than four SCT hits, due to the limited lever-arm of the pixel tracklet.
No significant dependence of the $q/\pt$ resolution on $\pt$ is observed.
The smearing procedure is validated in $Z \rightarrow \mu\mu$ events, as shown
in figure~\ref{figure:smearingFunction_fitting}(b). To validate the procedure, the smearing function is extracted using only one muon per $Z \rightarrow \mu\mu$ event, and the other muon is used to make the \pt spectrum from the re-fitted pixel tracklets. 
The \pt spectrum from the re-fitted pixel tracklets is compared to the one created
by convolving the track \pt spectrum with the smearing function.
The two spectra agree very well up to 12.5~\TeV{}.

\begin{figure}[t]
  \centering
  \subfigure[]{
    \includegraphics[width=0.46\textwidth]{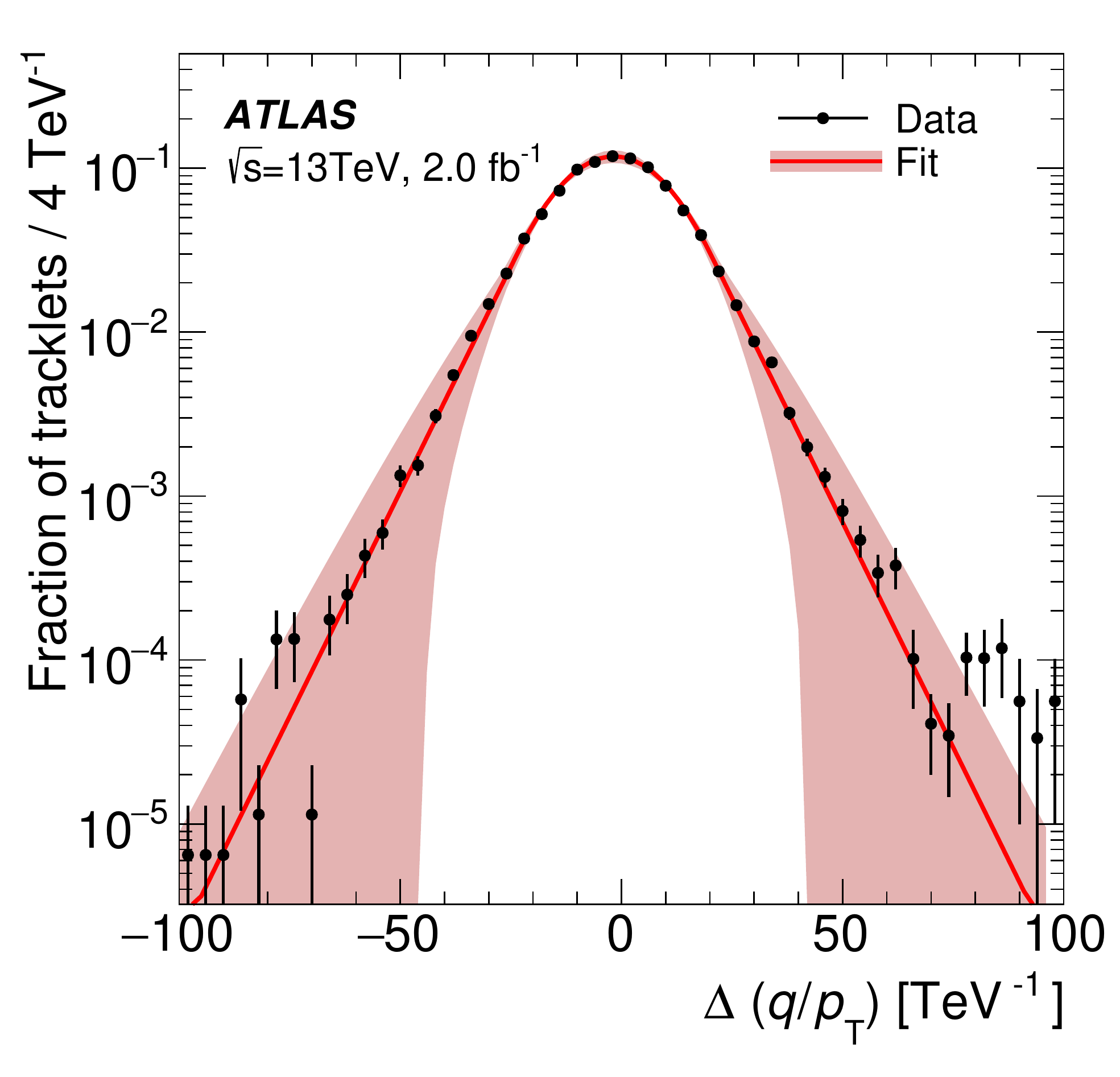}
  }
  \subfigure[]{
    \includegraphics[width=0.44\textwidth]{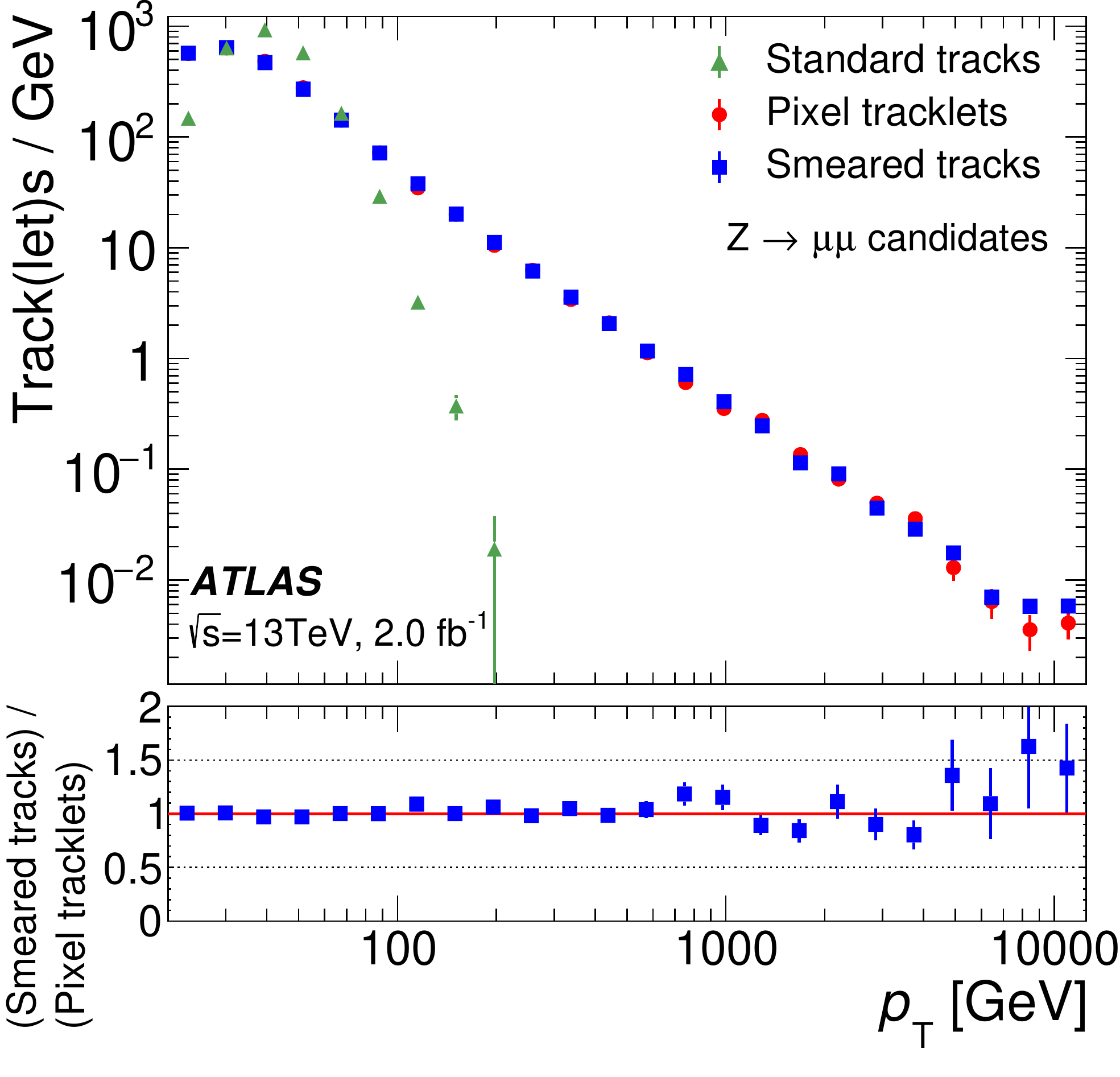}
  }
   \caption{(a) Distribution of the difference between $q/\pt$ of a pixel
     tracklet and a track in $Z \rightarrow \mu\mu$ events in
     data. The solid curve shows the smearing function (eq.~(\ref{eq:CrystalballLikeFunction})) 
     used to construct the background \pt template, which is described in
     section~\ref{sec:background}.
     The parameter values of the curve are
     $\alpha=$\num[round-precision=2]{1.66707e0},
     $\beta=$\num[round-precision=2]{-1.72142e+00} $\tev ^{-1}$ and
     $\sigma=$\num[round-precision=1]{13.2009} $\tev ^{-1}$.
     The red band indicates a 1$\sigma$ variation of the
     systematic uncertainty (see section~\ref{sec:uncertainties}). The data are normalised to unit area.
(b) Validation of the smearing procedure in $Z \rightarrow \mu\mu$ events in data.
The green and red points show the \pt distributions of tracks
and pixel tracklets respectively. The blue point shows the \pt spectrum obtained by
convolving the track \pt distribution with the smearing function.
The lower plot shows the ratio of the smeared
spectrum to the distribution of the pixel tracklets.} 
  \label{figure:smearingFunction_fitting}
\end{figure}

\subsubsection{Hadron background}\label{sec:hadronbg}

Assuming that the \pt spectrum of hadrons scattered in the ID is the same as that of
non-scattered hadrons, the \pt spectra of scattered hadrons can be extracted from
tracks in control samples of non-scattered hadrons. This assumption was verified with simulation.
The control samples are obtained by applying the same kinematic requirements as in the signal
regions and then selecting samples of tracks which satisfy the following requirements:
\begin{itemize}
\item The number of associated hits in the TRT must be larger than 15,
  and the number of associated hits in the SCT must be larger than 6.
\item There must be associated energy deposits in the calorimeter:
the transverse energy deposited in the calorimeter in a cone of $\Delta R=0.2$ around the
track, excluding the energy cluster associated with the track, ($\ET^{\mathrm{cone20}}$)
must satisfy $\ET^{\mathrm{cone20}} > 3$~\GeV{}, and the sum of cluster energies in a
cone of $\Delta R = 0.4$ around the track ($\sum_{\Delta R<0.4} \ET^{\mathrm{clus}}$)
divided by the \pt of the track must satisfy ${\sum_{\Delta R<0.4}
\ET^{\mathrm{clus}}/\pt>0.5}$.
\end{itemize}
The first requirement selects good-quality tracks which have not undergone scattering in the silicon layers. The second requirement removes electron and muon tracks from the control region. The \pt spectra of the control samples are convolved with the smearing function to take into account the resolution of the pixel tracklets.
Separate \pt spectra are prepared for the high-\met region and the
low-\met region in each channel.

\subsubsection{Charged-lepton background} \label{sec:lepton}

In order to obtain the \pt spectra of background tracklets originating from leptons, events containing a lepton without significant scattering due to hard bremsstrahlung are used.
The lepton \pt spectra obtained from these events are scaled to take into account the probability of significant scattering and are smeared to take into account the poor \pt resolution of pixel tracklets.
Events containing exactly one lepton which satisfy the same kinematic
requirements as for the signal regions, excluding the lepton-veto, are used.
The lepton is required to have an associated 
inner detector track 
with $\pt >16$ ~\GeV{}
which satisfies the same quality selection as for tracklets, except for the SCT veto and the isolation from
candidate electrons and muons.

The $\pt$ distribution of background tracklets from leptons is obtained by multiplying the
$\pt$ distribution of the lepton control sample by a transfer factor, which rescales the number of
 identified leptons to that of pixel tracklets. The transfer factor is found to be $\pt$-dependent for
 electrons, and $\eta$- and $\phi$-dependent for muons, as described below.

The transfer factor is extracted with a tag-and-probe method using $Z
\to \ell\ell$ events in data which are selected by a single-lepton trigger.
Tag and probe leptons are selected by applying requirements discussed below.
The tag--probe pair is further required to have an invariant mass within 10~\GeV{} of the $Z$ boson pole mass.

A tag electron is required to fully satisfy track-based isolation criteria and likelihood-based `Tight' electron identification criteria, to match the
electron which triggered the event and to have $\pt>30$~\GeV. 
Probe electrons are
identified as clusters of energy in the calorimeter with an associated track satisfying
the quality, isolation, high-\pt and geometrical acceptance requirements
defined in section~\ref{sec:object} for signal tracks and tracklets. The probe track has to satisfy either the full pixel tracklet selection, including the disappearance condition, or the tight electron selection.
The transfer factor is defined as the ratio of the number of probe
electrons which satisfy the
full tracklet selection to the number of probe electrons which satisfy the tight electron selection, as a function of
electron $\pt$. The transfer factor is O($10^{-2}$)--O($10^{-4}$), depending on electron $\pt$,
and is below $10^{-5}$ for electrons with $\pt > 50~\GeV{}$.

A muon used as a tag must satisfy track-based isolation criteria and cut-based `Tight' identification criteria.
The transfer factor for muons is the product of two components: the probability for a muon ID track
to be classified as disappearing and the probability for a muon ID track not to have
an associated MS track. As the pixel tracklets in the signal region are required to be isolated from MS tracks, the second component of the transfer factor allows an estimation of the expected normalisation as well as the $\pt$ distribution of the muon background. The first component of the muon transfer factor is estimated with a
method similar to that used for the electron transfer factor. The same selection
criteria as the electron case are applied to the tag and probe muons, replacing the electron
identification criteria with those for the muon. The first component of the muon transfer
factor is found to be $4.5\times10^{-4}$. The second component is necessary because an MS track is used
as a probe to measure the first component. The second component is evaluated with a similar tag-and-probe method,
where the probe is taken from a sample of well-measured tracks which pass through the full ID, selected by requiring more than 15 TRT hits on the track. The
probability for an MS track to be geometrically matched to an ID track is calculated from this sample. 
The transfer factor is measured as a function of $\eta$ and $\phi$ to fully take into
account the detector geometry. The second component of the transfer factor for muons is found to be of the order of $10^{-2}$ to $10^{-1}$.
The \pt spectra of the lepton control samples are scaled by the transfer factors, then
convolved with the smearing function.
Two different \pt spectra are prepared, one for the high-\met region and one for the low-\met region, while keeping the same requirements as in the signal region.
The expected numbers of muon background events in the low-\met region and in the high-\met region are estimated by scaling the number of events in the muon control samples by the transfer factor.

\subsubsection{Templates for scattered particles} \label{sec:scatteredparticles}
The control samples for hadron and electron components are found to
have similar \pt distributions, which is due in part to the fact that the isolation requirement for tracklets to be separated from jets affects both the electron and hadron background similarly. The two components are therefore combined in the fitting. The muon component is treated separately as the muon control samples are found to have a different \pt distribution.

\subsubsection{Fake tracklets} \label{sec:fakes}
Fake tracklets are formed from a random combination of hits.
The $d_0$ distribution of fake tracklets is broad, whereas the
high-$\pt$ chargino tracklets have good impact parameter resolution and
therefore have values of $d_0$ which cluster around zero.
The fake-tracklet control region is defined by requiring
$|d_0|/\sigma(d_0) >10$, and by removing the $\met$ requirement. This region is dominated by fake tracklets.
The \pt spectra of fake tracklets are modelled with the following empirical functional form:
\begin{equation}
f\left(\pt\right) = \exp\left(- p_0 \cdot \log(\pt) - p_1 \cdot (\log(\pt))^2\right),
\label{fakefunction}
\end{equation}
\noindent where $p_0$ and $p_1$
are fit parameters.
Figure \ref{figure:CR_pt} shows the \pt distribution of pixel tracklets in the fake-tracklet control
region along with a histogram filled from the result of the fit.
The \pt spectrum shape is confirmed to be independent of $\met$ by comparing it in three $\met$ regions: 
$\met<90~\GeV$, $90~\GeV<\met<140~\GeV$ and $\met>140~\GeV$.
A small dependence of the fit parameters on $|d_0|/\sigma(d_0)$ is
observed by comparing the parameters obtained in three regions: 
$10 < |d_0|/\sigma(d_0) <20$, $20< |d_0|/\sigma(d_0) <30$ and $30< |d_0|/\sigma(d_0) <100$. The size of the dependence on $|d_0|/\sigma(d_0)$ is added as an uncertainty in the \pt template shape.

\begin{figure}[t]
  \centering
  \subfigure[]{
    \includegraphics[width=0.48\textwidth]{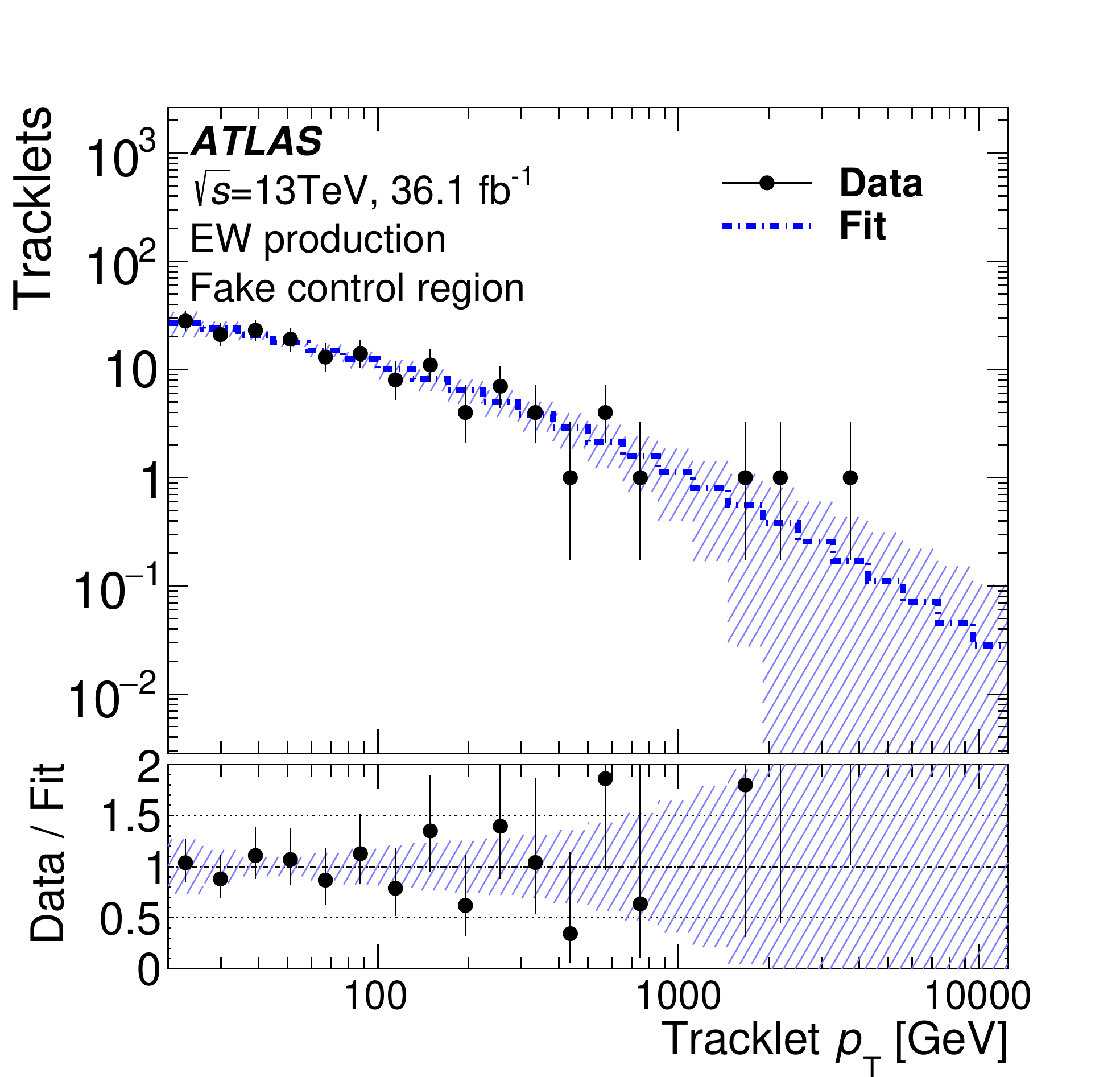}
  }
  \subfigure[]{
    \includegraphics[width=0.48\textwidth]{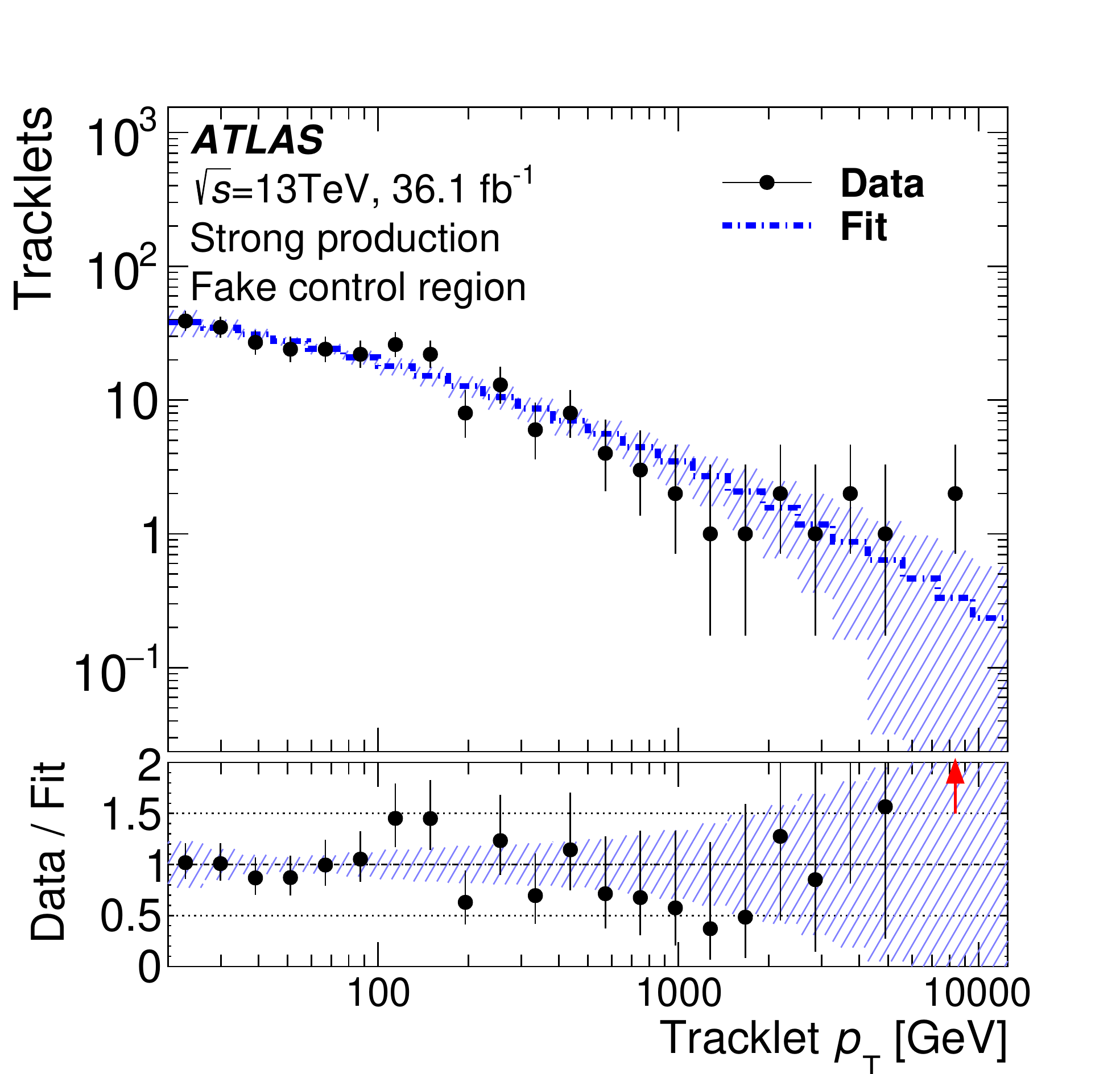}
  }
  \caption{Fit on the fake-tracklet control sample for (a) the electroweak production 
    channel and (b) the strong production channel. The black markers show data. The blue line and the
     band show the histogram made from the fit function and its uncertainty.
The bottom plot shows the ratio of the observed data to the fit histogram.
The chi-square per degrees of freedom of the fit are $5.4/14$ and $8.5/19$ for the electroweak and strong production channels respectively.
Red arrows in the Data/Fit ratio indicate bins where the corresponding entry falls outside the plotted range.}
  \label{figure:CR_pt}
\end{figure}

\subsection{Signal templates} \label{sec:signalpt}

The signal \pt spectrum is estimated by smearing the generator-level \pt distribution of
charginos in the signal simulation for each signal parameter
point.
The smearing function parameters are determined from muons in data,
but shifted by the differences between the parameter values
found for charginos and muons in simulation due to the difference between their masses. This smearing is performed because the tracklet $\pt$ resolution measured in reconstructed simulated samples is narrower than the resolution measured in data.
\subsection{Fit to the \pt spectrum} \label{sec:fit}

The extended likelihood function, described in detail in appendix~\ref{likelihood}, consists of signal and background components.
The background components represent tracklets from muons, fakes, and the sum of hadron and electron contributions.
The fit parameters are the signal strength, the normalisations of the sum
of the hadron and electron, muon, and fake-tracklet backgrounds, the fake-tracklet $\pt$ distribution's fit parameters, and nuisance parameters.
The nuisance parameters are allowed to float in the fit with Gaussian constraints to include systematic uncertainties, discussed in section \ref{sec:uncertainties}.
The number of signal events and the sum of hadron and electron events are fit without a Gaussian-constraint term.
The number of muon events and the sum of hadron and electron events are fit with independent individual normalisation factors in the low-\met and high-\met regions. The number of muon events is constrained by a Gaussian term which represents the expectation described in section~\ref{sec:lepton}. The statistical uncertainty in the transfer factors for muons is propagated to the final template.
The fake-tracklet control region is divided into two parts, a low-\met
and a high-\met
fake-tracklet control region, by applying the same \met requirement as in the signal region.
The signal regions and the two parts in the fake-tracklet control region are fit simultaneously and the ratio of the number of fake tracklets in the high-\met signal region to that in the low-\met region is constrained to the same value as in the fake-tracklet control region.

\section{Systematic uncertainties}\label{sec:uncertainties}

\subsection{Background uncertainties} \label{sec:background_uncertainty}

An uncertainty in the shape of the hadron and electron \pt template was estimated as the
maximum difference between the hadron and electron individual templates and found to be
negligible. As a combined template is used for hadrons and electrons, the difference
in tracklet $q/\pt$ resolutions between hadrons and electrons in simulation is added to the systematic uncertainty in the smearing function
for the combined template.
The red band in figure~\ref{figure:smearingFunction_fitting} shows the uncertainty in the
smearing function.

Possible differences between the signal and the fake-tracklet control region leading to systematic
uncertainties in the shape of the \pt spectrum of the fake-tracklet
background are taken into account.
The uncertainty is estimated from the $d_0$ significance dependence of the parameters of
the fake-tracklet \pt spectrum function defined in eq.~(\ref{fakefunction}) for the fake-tracklet control region.
A conservative uncertainty of 100\% is assigned to the ratio of the number of fake tracklets in the low-\met control region to the number in the high-\met control region.

\subsection{Signal uncertainties} \label{sec:signal_uncertainty}

A breakdown of the systematic uncertainties in the expected number of signal events passing the signal region requirements is shown in table~\ref{table:summary_uncertainty}. In addition, an uncertainty in the $\pt$ spectrum shape, due the uncertainty in the $\pt$ resolution, is taken into account.

High-$\pt$ jets originating from ISR and final state radiation (FSR) alter the signal acceptance.
Uncertainties in the modelling of ISR and FSR are estimated by varying the renormalisation, factorisation and merging scales from 0.5 to 2 times their nominal values, and by comparing samples with one and two additional partons in the matrix
element with {\scshape MG5\_}a{\scshape MC@NLO}+{\scshape Pythia8}.
For the strong channel, the ISR/FSR uncertainty is small when the mass difference between
the gluino and chargino is large; however, the uncertainty grows to about 10\% when the mass
difference is smaller than 200~\GeV, as signal events start to be rejected by the requirement on the jet \pt.
The uncertainties in the jet energy scale and resolution are estimated by the techniques in refs.~\cite{ATL-PHYS-PUB-2015-015,PERF-2011-03,PERF-2011-05,ATLAS-CONF-2015-037,PERF-2011-04}.

The uncertainty in the trigger efficiency is small because it is measured from data, as described in
section~\ref{sec:samples}. For the signal \pt resolution, a conservative uncertainty, corresponding to 100\% of the effect of multiple
scattering, is added to the uncertainty in the values of parameters in the $q/\pt$ smearing function.

The pile-up modelling uncertainty is estimated by varying the number of collisions per
bunch crossing in simulation by its uncertainty of 10\% of the nominal value. The signal
reconstruction efficiency decreases as the number of pile-up interactions increases because it becomes more likely for pixel-detector hits originating from charginos to be used by tracks from other particles.

The uncertainty in the chargino reconstruction efficiency (tracklet efficiency) can be split into four
components: (1) the uncertainty in the probability for a chargino to produce a set of pixel-detector hits which
can satisfy the tracklet quality selection, (2)
the uncertainty in the efficiency to reconstruct a tracklet when a chargino leaves a set of good hits
which satisfies the tracklet quality selection, (3) the uncertainty in the track
reconstruction efficiency, which depends on the number of pile-up interactions, (4) the uncertainty in the $d_0$
significance selection.
The first two components are estimated using $Z \rightarrow \mu\mu$ events, which are selected
with the same requirements as for the data sample used to estimate the smearing function.
The tracklet data-quality selection requirements are applied to the muon tracks in the sample.
The first component is estimated from the difference in the efficiency of these requirements between data and simulation.
The second component is estimated by re-fitting the muon tracks using only the pixel hits,
and comparing the tracklet reconstruction efficiencies in data and simulation.
The third component is included in the uncertainty in the pile-up modelling described
already.
The fourth component is estimated by shifting the measured $|d_0|/\sigma(d_0)$ distribution
by its uncertainty; the change in the $|d_0|/\sigma(d_0)$ selection efficiency is added
to the systematic uncertainty.

Theoretical uncertainties in the signal cross-section are estimated by computing the changes in the cross-section when the renormalisation and factorisation scales, the choice of PDFs and the strong coupling constant, $\alpha_\mathrm{S}$, are varied.
Renormalisation and factorisation scales are varied by factors of 0.5 and 2 from their nominal value.
The PDF uncertainty is estimated as the maximum of the uncertainty from the {\scshape CTEQ6.6}~\cite{Pumplin:2002vw} uncertainty band at 68\% confidence level and the difference between {\scshape CTEQ6.6} and {\scshape MSTW2008 NLO} PDF sets.
Each uncertainty is varied independently and their effects are added in quadrature.

\begin{table}[t]
  \centering
  \begin{tabular}{m{50mm}<{\hfill}S[round-mode=off]S[round-mode=off]}
    \hline
    \hline
Relative uncertainties [\%]
           & {Electroweak channel} & {Strong channel} \\
    \hline
    MC statistical uncertainty           & 6.6 & 6.5 \\
    ISR/FSR                            & 7.6 & 0.2 \\
    Jet energy scale and resolution    & 2.0 & 0.7 \\
    Trigger efficiency                 & 0.2 & <0.1 \\
    \cline{2-3}
    Pile-up modelling                  & \multicolumn{2}{@{\hskip -0.65em}c}{11} \\
    Tracklet efficiency                & \multicolumn{2}{c}{6.9} \\
    Luminosity                         & \multicolumn{2}{c}{3.2} \\
    \cline{2-3}
    {\bfseries Sub-total}              & 17 & 15 \\
    \hline
    Cross-section  & 6.4 & 28 \\
    \hline
    {\bfseries Total}                  & 18 & 32 \\
    \hline
    \hline
  \end{tabular}
  \caption{Systematic uncertainties in the signal event yields at
     $m_{\chinoonepm} = 400$~\GeV\ for the electroweak channel
      and at $m_{\gluino} = 1600$~\GeV, $m_{\chinoonepm} = 500$~\GeV\ for the strong channel. The lifetime of the chargino is not relevant here.
      The uncertainty in the cross-section of the strong production is large due to the
      large effect from the PDF uncertainty.
  }
  \label{table:summary_uncertainty}
\end{table}

\section{Results and interpretation}\label{sec:results}

The tracklet \pt spectra are shown in figure~\ref{figure:fit}, along with the results of the fit to the background-only hypothesis.
The observed \pt distributions are well described by the background predictions in
the low-$\met$ regions.
When fitting to the background+signal hypothesis, no significant excess above the expected SM processes is found at high tracklet $\pt$ in high-$\met$ regions.
Model-dependent upper limits on the signal strength are computed using the profile-likelihood ratio
\cite{Cowan:2010js} as a test statistic and using the asymptotic
formula in ref.~\cite{Cowan:2010js}, fitting the \pt spectrum in the
full range. The confidence levels are computed by following the CL${}_\mathrm{s}$ prescription~\cite{Read:2002hq}.
Upper limits on the number of signal events are converted into limits on the visible cross-section ($\sigma_{\mathrm{vis}}^{95\%}$) of signal processes by dividing by the integrated luminosity of the data.

\begin{figure}[t]
  \centering
  \subfigure[Electroweak channel low-$\met$ region]{
    \includegraphics[width=0.45\linewidth]{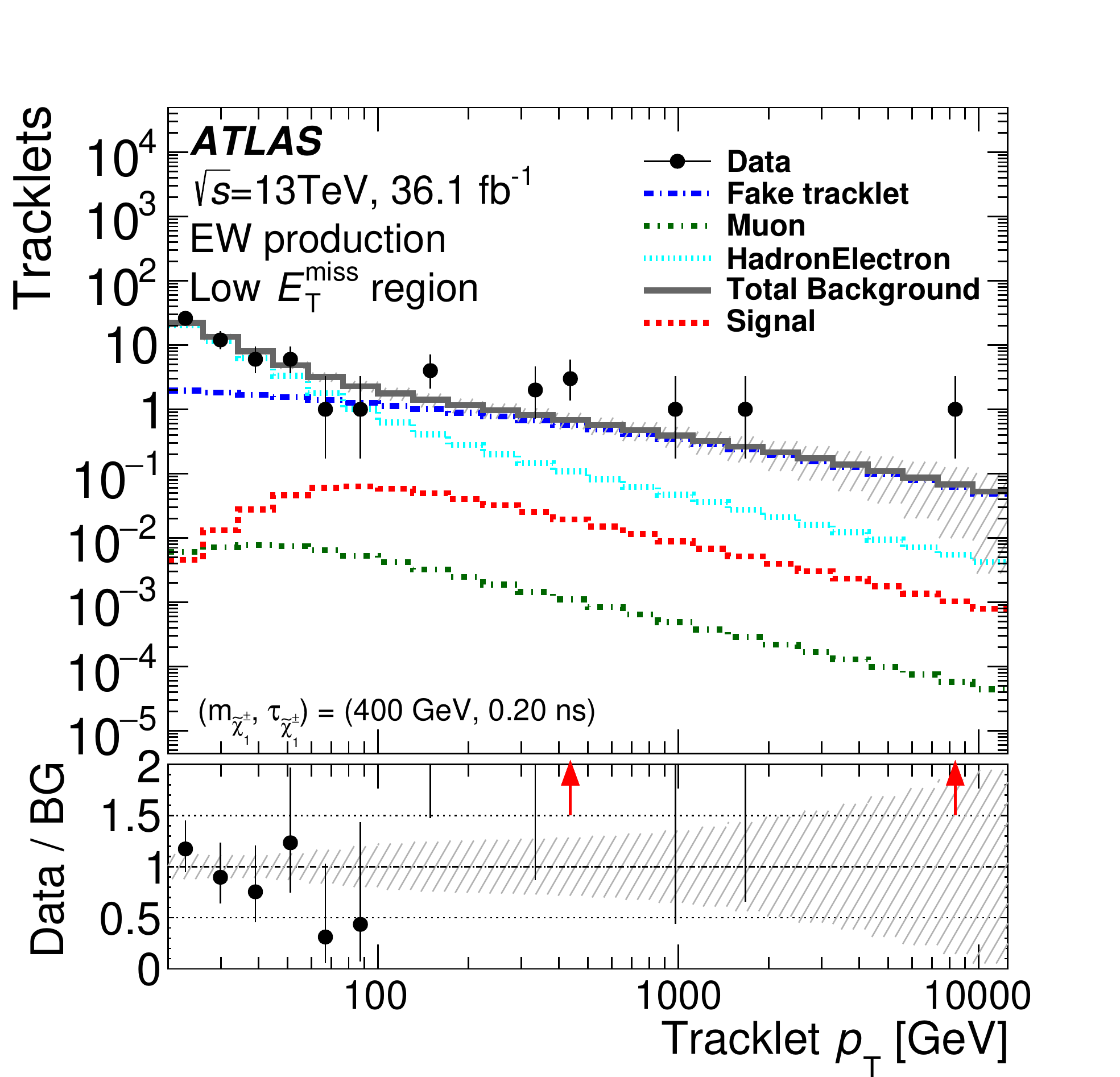}
  }
  \subfigure[Strong channel low-$\met$ region]{
    \includegraphics[width=0.45\linewidth]{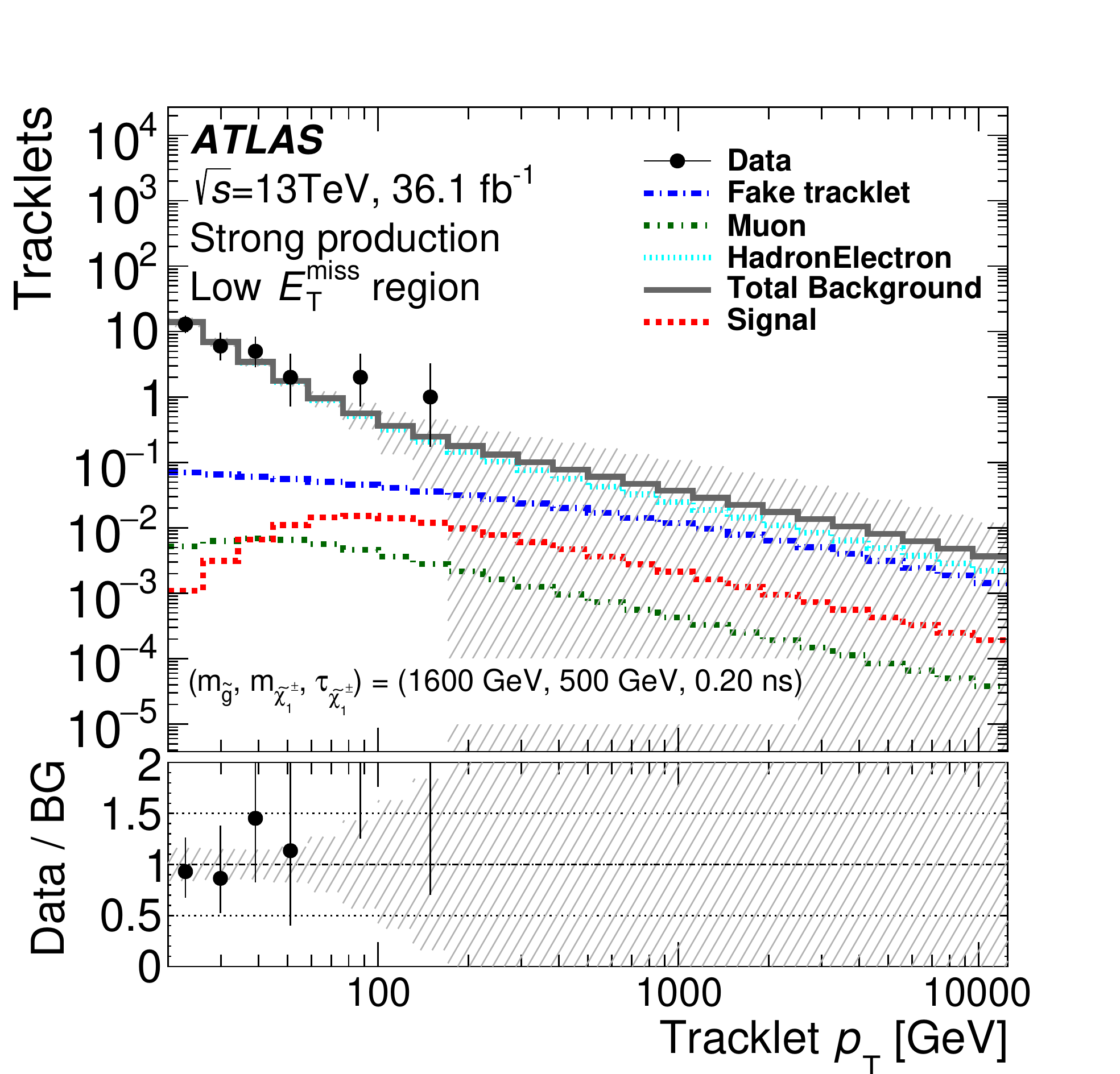}
  }
  \subfigure[Electroweak channel high-$\met$ region]{
    \includegraphics[width=0.45\linewidth]{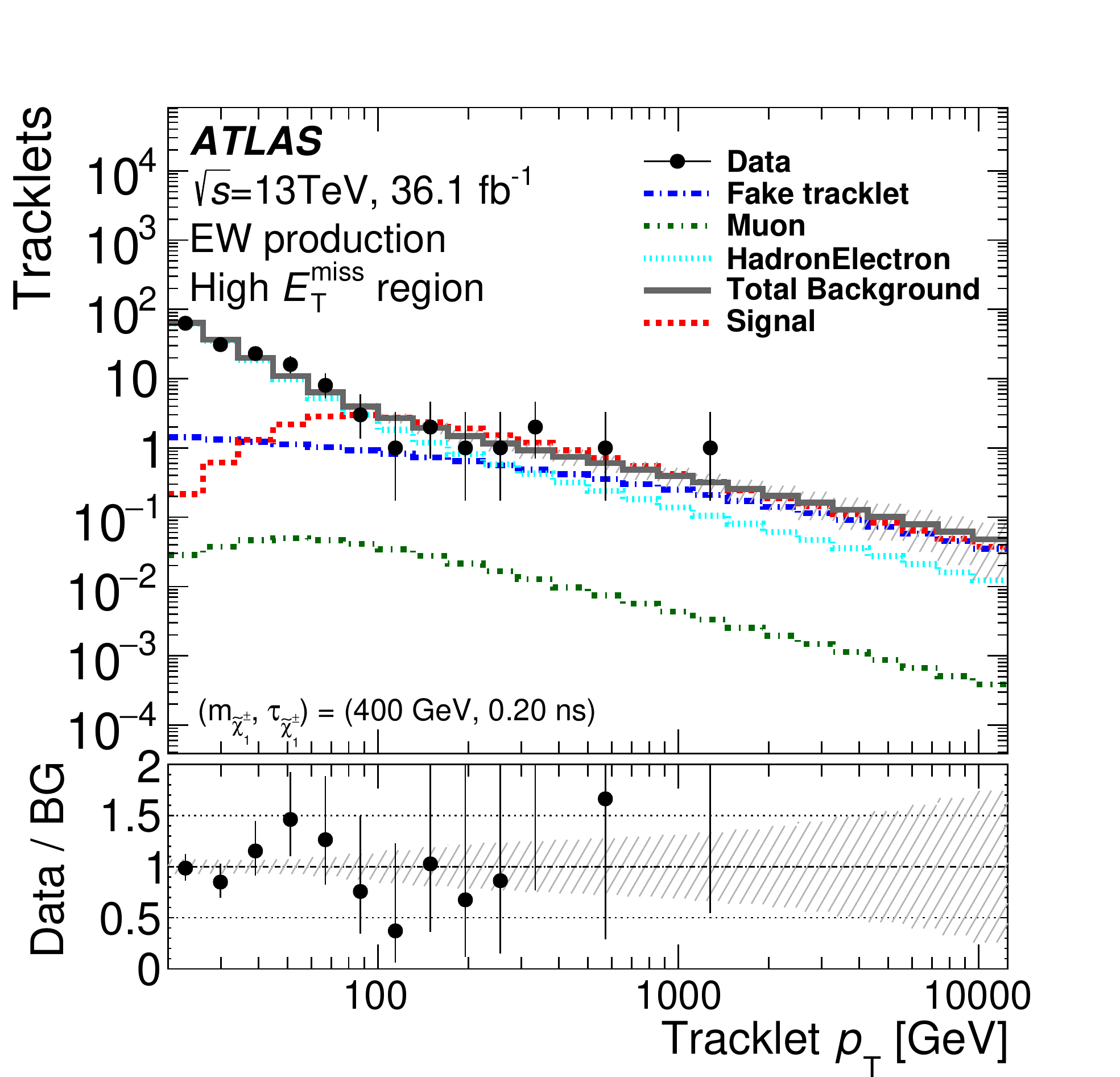}
  }
  \subfigure[Strong channel high-$\met$ region]{
    \includegraphics[width=0.45\linewidth]{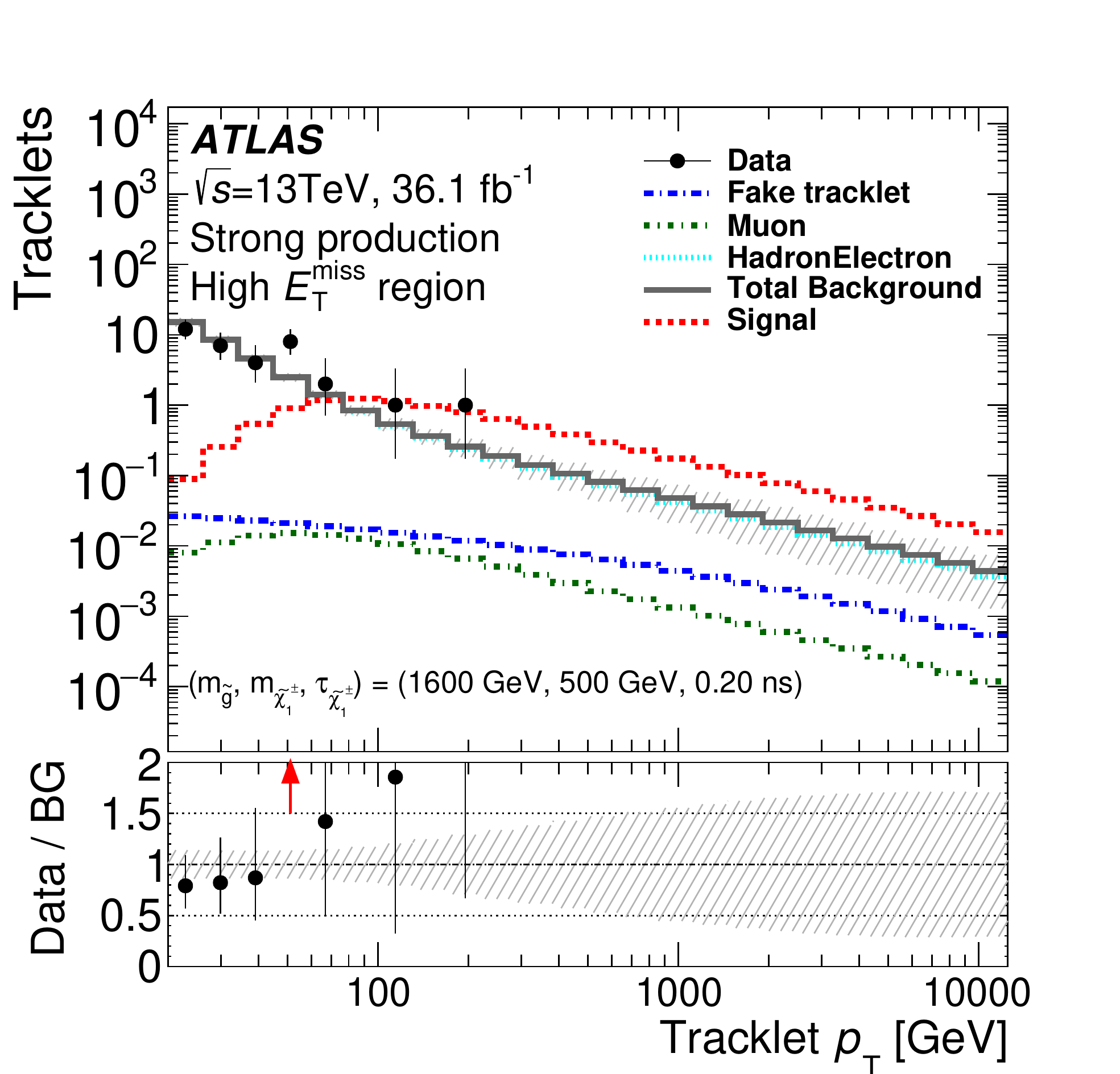}
  }
  \caption{Pixel-tracklet \pt spectrum in various regions: (a)
    electroweak channel in the low-$\met$ region, (b) strong channel
    in the low-$\met$ region, (c) electroweak channel in the
    high-$\met$ region, and (d) strong channel in the high-$\met$ region. Observed data are shown with
     markers and the background components for the
     background-only fit are shown with lines.
     In the strong channel, total background lines overlap hadron and electron background lines.
     An example of the expected signal spectrum
at $\tau_{\chinoonepm}=0.2$~ns and $m_{\chinoonepm}=400$~\GeV\ for the electroweak channel and $m_{\gluino}=1600$~\GeV, $m_{\chinoonepm}=500$~\GeV\ for the strong channel is overlaid for comparison. The bottom panels show the ratio of the data to the background predictions.
     The error band shows the uncertainty in the background prediction including both the 
     statistical and systematic uncertainties.
Red arrows in the Data/BG ratio indicate bins where the corresponding entry falls outside the plotted range.}
  \label{figure:fit}
\end{figure}

Model-independent limits are calculated from the expected and observed
event yields in the region where the tracklet \pt is above 100~\gev. 
Table~\ref{table:pvalue} lists the observed event yields, expected backgrounds, expected signal
yields and model-independent upper limits on the visible signal cross-section in the high-$\met$ region.

\begin{table*}[t]
  \centering
  \begin{tabular}{lS[separate-uncertainty,table-format=4.4(4)]S[separate-uncertainty,table-format=4.4(4)]}
    \hline
    \hline
                  & {Electroweak channel}    & {Strong channel}  \\
    \hline
    \multicolumn{3}{c}{\bfseries{Number of observed events with $\pt > 100$~\GeV{} in high-\met\ regions}} \\
                                        & 9             & 2                \\
    \hline
    \multicolumn{3}{c}{\bfseries{Number of expected events with $\pt > 100$~\GeV{} in high-\met\ regions}} \\
    Hadron+electron background & 6.1   (6)  & 1.78   (32) \\
    Muon background            & 0.15  (9)  & 0.05   (8)  \\
    Fake background            & 5.5   (33) & 0.1    (4)  \\
    Total background           & 11.8  (31) & 1.9    (4)  \\
    \hline
    $p_0$                                               & 0.5         & 0.47      \\
    Observed $\sigma_{\mathrm{vis}}^{95\%}~\mathrm{[fb]}$  & 0.22            & 0.12             \\
    Expected $\sigma_{\mathrm{vis}}^{95\%}~\mathrm{[fb]}$  & {\hspace{-1.28em}$0.28^{+0.11}_{-0.08}$} & {\hspace{-1.28em}$0.12^{+0.07}_{-0.04}$} \\
    \hline
    \multicolumn{3}{c}{\bfseries{Number of expected signal events with $\pt > 100$~\GeV{} in high-\met\ regions}} \\
                                    &  13.5 (21)  & 5.6 (8)    \\
    \hline
    \hline
  \end{tabular}
  \caption{Observed events, expected background for null signal, and 
 expected signal yields for two benchmark models: electroweak channel with {\small
    ($m_{\chinoonepm}$, $\tau_{\chinoonepm}$) = (400~\gev{}, 0.2~ns)} and strong channel with {\small ($m_{\gluino}$, $m_{\chinoonepm}$, $\tau_{\chinoonepm}$) = (1600~\gev{}, 500~\gev, 0.2~ns)}
 in the high-$\met$ region.
Also shown are the probability of a background-only experiment being more signal-like than
observed ($p_0$) and the upper limit on the model-independent visible cross-section at 95\% CL.
The uncertainty in the total background yield is different from the sum of uncertainties in quadrature due to anti-correlation between different backgrounds.}
  \label{table:pvalue}
\end{table*}

Figure~\ref{figure:exclusionew} shows the model-dependent exclusion limits in the
($m_{\chinoonepm},\tau_{\chinoonepm}$) plane for the electroweak channel, where
$\tau_{\chinoonepm}$ is the lifetime of the chargino.
A large region is excluded by this analysis while the 8~\TeV{} result \cite{SUSY-2013-01} has higher
sensitivity for long lifetimes due to the use of longer tracklets.
For $\tau_{\chinoonepm}\sim 0.2$~ns, which corresponds to $\mdiff\sim 160~\MeV${}
in the pure wino LSP model, winos with a mass up to 460~\GeV\ are excluded at 95\% CL.
Figure~\ref{figure:exclusionstrong} shows the model-dependent exclusion limits in the 
$m_{\gluino}$--$m_{\chinoonepm}$ plane for the strong channel.
For a chargino lifetime of 0.2~ns, gluino masses up to 1.65~\TeV\ are excluded assuming a chargino mass of 460~\GeV, and chargino masses up to 1.05~\TeV\ are excluded assuming very compressed spectra with a mass difference between the gluino and the chargino of less than 200~\GeV.
Charginos are assumed to decay into a pion and a neutralino in the considered models.
However, the results
do not depend on this decay mode since the decay products of charginos cannot be detected due
to their low momentum.

\begin{figure}[t]
  \centering
  \includegraphics[width=0.8\linewidth]{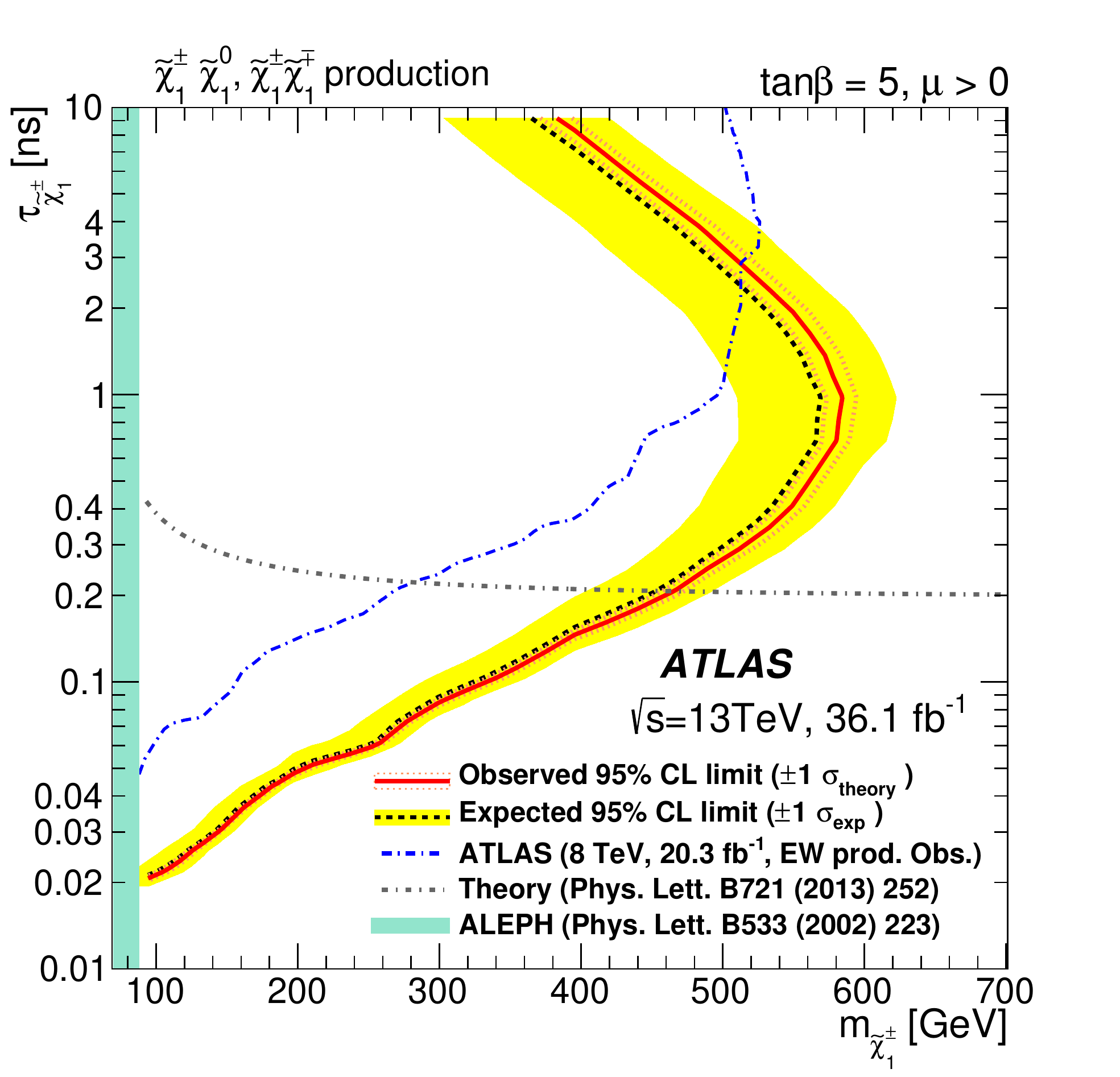}
  \caption{Exclusion limit at 95\% CL obtained in the electroweak production channel in terms
     of the chargino lifetime ($\tau_{\chinoonepm}$) and mass ($m_{\chinoonepm}$). The yellow band shows the 1$\sigma$ region of the
        distribution of the expected limits. The median of the expected limits is shown by a
        dashed line. 
        The red line shows the observed limit and the orange dotted lines around it show the impact on the observed limit of the variation of the nominal signal cross-section by $\pm1\sigma$ of its theoretical uncertainties.
      Results are compared with the observed limits obtained by the previous
      ATLAS search with disappearing tracks and tracklets~\cite{SUSY-2013-01} and an example of the
      limit obtained at LEP2 by the ALEPH experiment~\cite{Heister:2002mn}. The chargino lifetime as a function of the chargino mass is shown in the almost pure wino LSP scenario at the two-loop level~\cite{Ibe:2012hu}. }
  \label{figure:exclusionew}
\end{figure}

\begin{figure}[t]
  \centering
  \subfigure[]{
    \includegraphics[width=0.45\linewidth]{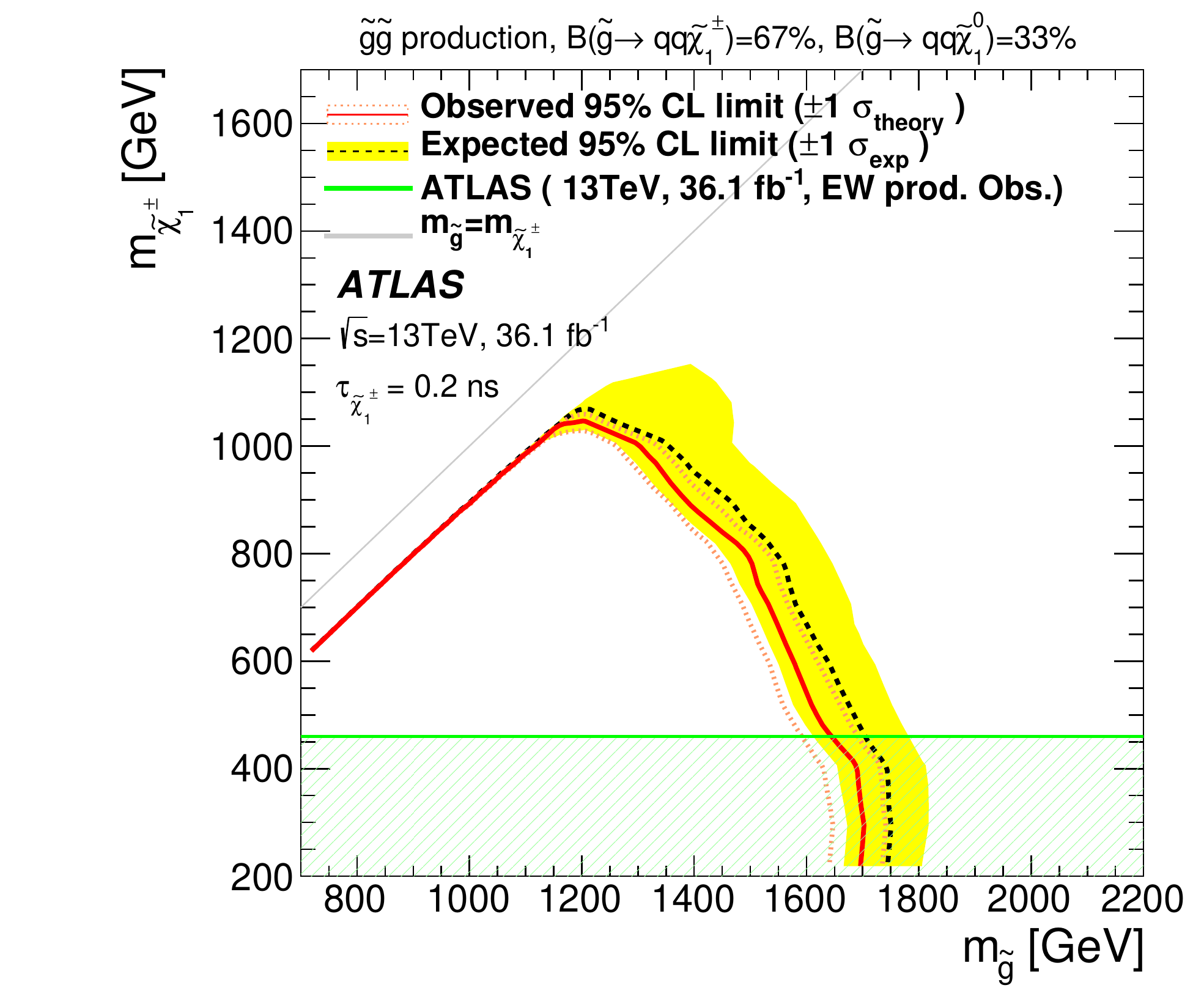}
  }
  \subfigure[]{
    \includegraphics[width=0.45\linewidth]{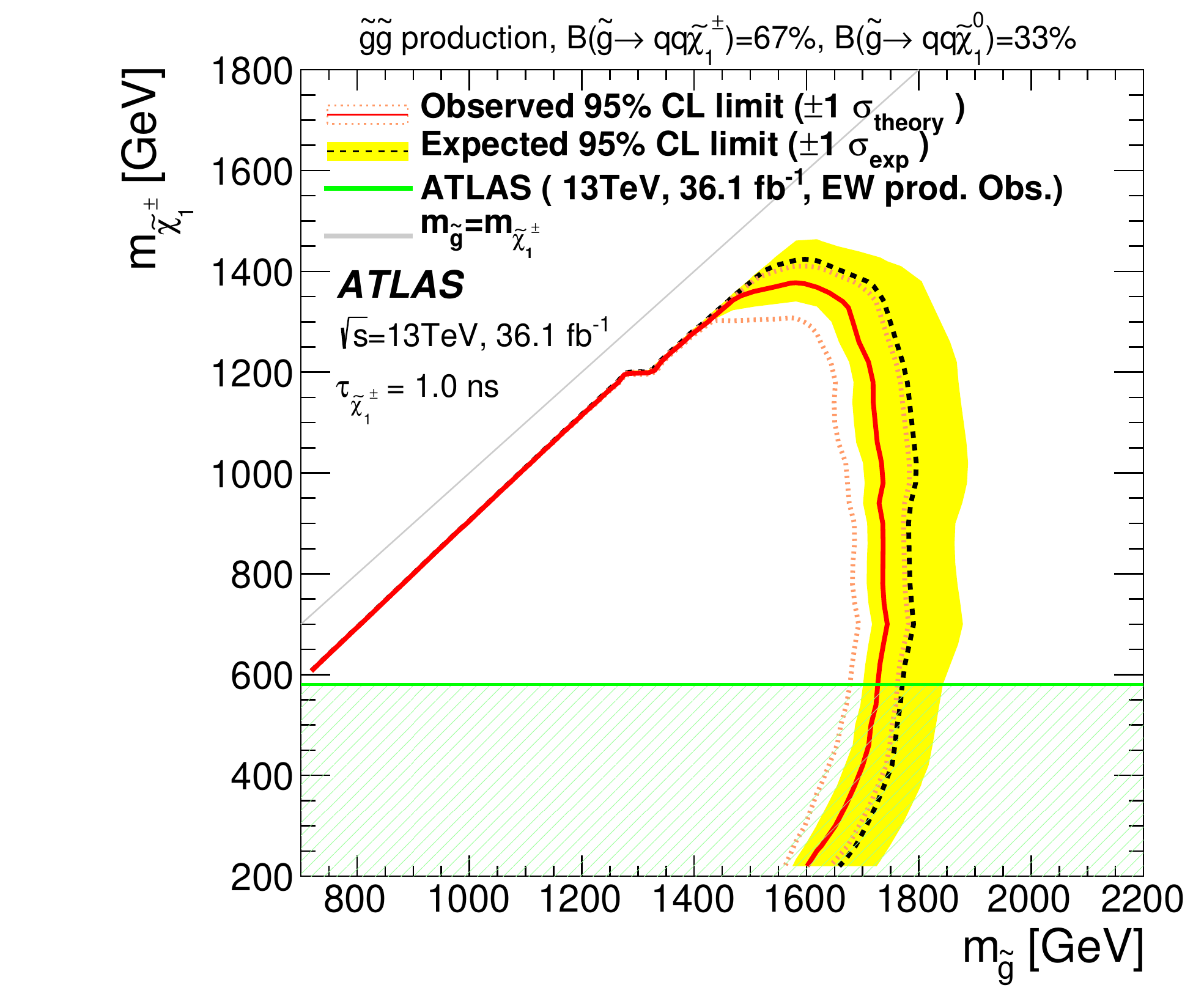}
  }

  \caption{Exclusion limit at 95\% CL obtained in the strong production channel in terms
     of the gluino and chargino masses. The limit is shown assuming a chargino lifetime of
        (a) 0.2~ns and (b) 1.0~ns. The yellow band shows the 1$\sigma$ region of the
        distribution of the expected limits. The median of the expected limits is shown by a
        dashed line. The red line shows the observed limit and the orange dotted lines around it show the impact on the observed limit of the variation of the nominal signal cross-section by $\pm1\sigma$ of its theoretical uncertainties.
        Observed limits in the electroweak production search are shown as a green shaded region.}
  \label{figure:exclusionstrong}
\end{figure}

The effects of systematic uncertainties are estimated using the exclusion significance,
which is defined as the number of standard deviations corresponding to the signal confidence CL${}_\mathrm{s}$.
Relative changes in the exclusion significance, when nuisance parameters are shifted by one standard deviation from their nominal values, are summarised in table \ref{table:uncertainty_effects}.
When shifting a parameter, the other nuisance parameters are fixed at their nominal values.

\begin{table}[t]
  \centering
  \begin{tabular}{l|S[round-mode=off]S[round-mode=off]}
    \hline
    \hline
Parameter
           & {Electroweak channel [\%]} & {Strong channel [\%]} \\
    \hline
    Expected signal events                         & 11 & 13 \\
    $\alpha$ in signal \pt resolution function     & 0.8  & 1.5  \\
    $\sigma$ in signal \pt resolution function     & 5.3  & 7.2  \\
    $\log r_{\mathrm{ABCD}}$                       & 15 & <0.1 \\
    $\alpha$ in background \pt resolution function & 5.0  & 1.2  \\
    $\sigma$ in background \pt resolution function & 2.2  & 5.0  \\
    $p_0$ parameter of the fake-BG \pt function    & 2.5  & <0.1 \\
    $p_1$ parameter of the fake-BG \pt function    & 8.5  & 0.1  \\
    Expected number of muon events                 & 0.5  & 0.9  \\
    \hline
    \hline
  \end{tabular}
  \caption{Effects of systematic uncertainties on the signal exclusion significance at
     $m_{\chinoonepm}=400$~\GeV\ for the electroweak channel
      and at $m_{\gluino}=1600$~\GeV, $m_{\chinoonepm}=500$~\GeV\ for the strong channel. The lifetime of the chargino is not relevant here.
      Effects of uncertainties on the fake-tracklet background is smaller in the strong channel
      analysis because the estimated number of the fake-tracklet background events is small.
  }
  \label{table:uncertainty_effects}
\end{table}

\clearpage

\section{Conclusions}\label{sec:conclusion}

A new search for long-lived charginos yielding a pixel-tracklet signature was
performed based on $pp$ collision data collected by the ATLAS experiment at the LHC in 2015 and 2016 at $\sqrt{s} = 13$~\TeV{}, corresponding to an integrated luminosity of 36.1~$\ifb${}.
Tracklets with hits only in the pixel detector are used to improve the sensitivity for short chargino lifetimes.
The \pt distribution of the observed pixel tracklets is found to be consistent with the background prediction.
A lower limit on $m_{\chinoonepm}$ for electroweak production of long-lived charginos with
a proper lifetime of 0.2~ns, corresponding to a mass-splitting between the charged and neutral wino of 160~\MeV{}, in the pure wino LSP model is set at 460~\GeV{} at 95\% CL.
If charginos with a proper lifetime of 0.2~ns are produced in the decay cascade of pair-produced gluinos, 
gluino masses below 1.65~\TeV{} are excluded for a chargino mass of 460~\GeV{}, and chargino masses 
below 1.05~\TeV{} are excluded in the case of compressed spectra with a mass difference of 200~\GeV{} 
between the gluino and the chargino.

\section*{Acknowledgements}


We thank CERN for the very successful operation of the LHC, as well as the
support staff from our institutions without whom ATLAS could not be
operated efficiently.

We acknowledge the support of ANPCyT, Argentina; YerPhI, Armenia; ARC, Australia; BMWFW and FWF, Austria; ANAS, Azerbaijan; SSTC, Belarus; CNPq and FAPESP, Brazil; NSERC, NRC and CFI, Canada; CERN; CONICYT, Chile; CAS, MOST and NSFC, China; COLCIENCIAS, Colombia; MSMT CR, MPO CR and VSC CR, Czech Republic; DNRF and DNSRC, Denmark; IN2P3-CNRS, CEA-DRF/IRFU, France; SRNSFG, Georgia; BMBF, HGF, and MPG, Germany; GSRT, Greece; RGC, Hong Kong SAR, China; ISF, I-CORE and Benoziyo Center, Israel; INFN, Italy; MEXT and JSPS, Japan; CNRST, Morocco; NWO, Netherlands; RCN, Norway; MNiSW and NCN, Poland; FCT, Portugal; MNE/IFA, Romania; MES of Russia and NRC KI, Russian Federation; JINR; MESTD, Serbia; MSSR, Slovakia; ARRS and MIZ\v{S}, Slovenia; DST/NRF, South Africa; MINECO, Spain; SRC and Wallenberg Foundation, Sweden; SERI, SNSF and Cantons of Bern and Geneva, Switzerland; MOST, Taiwan; TAEK, Turkey; STFC, United Kingdom; DOE and NSF, United States of America. In addition, individual groups and members have received support from BCKDF, the Canada Council, CANARIE, CRC, Compute Canada, FQRNT, and the Ontario Innovation Trust, Canada; EPLANET, ERC, ERDF, FP7, Horizon 2020 and Marie Sk{\l}odowska-Curie Actions, European Union; Investissements d'Avenir Labex and Idex, ANR, R{\'e}gion Auvergne and Fondation Partager le Savoir, France; DFG and AvH Foundation, Germany; Herakleitos, Thales and Aristeia programmes co-financed by EU-ESF and the Greek NSRF; BSF, GIF and Minerva, Israel; BRF, Norway; CERCA Programme Generalitat de Catalunya, Generalitat Valenciana, Spain; the Royal Society and Leverhulme Trust, United Kingdom.

The crucial computing support from all WLCG partners is acknowledged gratefully, in particular from CERN, the ATLAS Tier-1 facilities at TRIUMF (Canada), NDGF (Denmark, Norway, Sweden), CC-IN2P3 (France), KIT/GridKA (Germany), INFN-CNAF (Italy), NL-T1 (Netherlands), PIC (Spain), ASGC (Taiwan), RAL (UK) and BNL (USA), the Tier-2 facilities worldwide and large non-WLCG resource providers. Major contributors of computing resources are listed in ref.~\cite{ATL-GEN-PUB-2016-002}.

%

 \clearpage
 \appendix
 \part*{Appendix}
\section{Likelihood function}\label{likelihood}
The likelihood function is:
\begin{eqnarray}
  \mathcal{L}_{\mathrm{Total}} &=& \mathcal{L}_{\mathrm{shape}} \times \mathcal{L}_{\mathrm{syst}} \times\mathcal{L}_{\mathrm{syst, fake}}; \\
  \mathcal{L}_{\mathrm{shape}} &=&
           \frac{{\mathrm{e}}^{-\left(n^H_{\mathrm{s}} + n^H_{\mathrm{h+e}} + n^H_{\mathrm{\mu}} + n^H_{\mathrm{f}}\right)}}{n^H_{\mathrm{obs}}!}\times\frac{\mathrm{e}^{-\left(n^L_{\mathrm{s}} + n^L_{\mathrm{h+e}} + n^L_{\mathrm{\mu}} + n^L_{\mathrm{f}}\right)}}{n^L_{\mathrm{obs}}!} \nonumber\\
  && \times \frac{{\mathrm{e}}^{-n^H_{\mathrm{Fake CR}}}}{n^H_{\mathrm{Fake CR, obs}}!}\times \left(n_{\mathrm{Fake CR}}^{H}\right)^{n^H_{\mathrm{Fake CR, obs}}}
     \times \frac{{\mathrm{e}}^{-n^L_{\mathrm{Fake CR}}}}{n^L_{\mathrm{Fake CR, obs}}!}\times \left(n_{\mathrm{Fake CR}}^{L}\right)^{n^L_{\mathrm{Fake CR, obs}}} \nonumber\\
  && \times      \prod^{n^H_{\mathrm{obs}}} \left(n^H_{\mathrm{s}}\mathcal{F}^H_{\mathrm{s}}\left(\pt;\sigma^{\mathrm{smearing}}_{\mathrm{s}},\alpha^{\mathrm{smearing}}_{\mathrm{s}}\right) + n^H_{\mathrm{h+e}}\mathcal{F}^H_{\mathrm{h+e}}\left(\pt;\sigma^{\mathrm{smearing}}_{\mathrm{h+e}},\alpha^{\mathrm{smearing}}_{\mathrm{h+e}}\right)\right.\nonumber\\
  &&  \;\;\;\;\; \left. + n^H_{\mathrm{\mu}}\mathcal{F}^H_{\mathrm{\mu}}\left(\pt\right) + n^H_{\mathrm{f}}\mathcal{F}_{\mathrm{f}}\left(\pt;p_0, p_1\right)\right)\nonumber\\
  && \times      \prod^{n^L_{\mathrm{obs}}} \left(n^L_{\mathrm{s}}\mathcal{F}^L_{\mathrm{s}}\left(\pt;\sigma^{\mathrm{smearing}}_{\mathrm{s}},\alpha^{\mathrm{smearing}}_{\mathrm{s}}\right) + n^L_{\mathrm{h+e}}\mathcal{F}^L_{\mathrm{h+e}}\left(\pt;\sigma^{\mathrm{smearing}}_{\mathrm{h+e}},\alpha^{\mathrm{smearing}}_{\mathrm{h+e}}\right)\right.\nonumber\\
  &&  \;\;\;\;\; \left. + n^L_{\mathrm{\mu}}\mathcal{F}^L_{\mathrm{\mu}}\left(\pt\right) + n^L_{\mathrm{f}}\mathcal{F}_{\mathrm{f}}\left(\pt;p_0, p_1\right)\right),\\
  \mathcal{L}_{\mathrm{syst}} &=& \mathcal{N}\left(\alpha_{\mathrm{s}}^{H};\overline{\alpha_{\mathrm{s}}^{H}},\Delta\alpha_{\mathrm{s}}^{H}\right) \times \mathcal{N}\left(n_{\mathrm{\mu}}^{H};\overline{n_{\mathrm{\mu}}^{H}},\Delta n_{\mathrm{\mu}}^{H}\right) \times \mathcal{N}\left(\alpha_{\mathrm{s}}^{L};\overline{\alpha_{\mathrm{s}}^{L}},\Delta\alpha_{\mathrm{s}}^{L}\right) \times \mathcal{N}\left(n_{\mathrm{\mu}}^{L};\overline{n_{\mathrm{\mu}}^{L}},\Delta n_{\mathrm{\mu}}^{L}\right) \nonumber \\
  && \times \mathcal{N}\left(\sigma_{\mathrm{h+e}}^{\mathrm{smearing}};\overline{\sigma_{\mathrm{h+e}}^{\mathrm{smearing}}},\Delta\sigma_{\mathrm{h+e}}^{\mathrm{smearing}}\right) \times \mathcal{N}\left(\alpha_{\mathrm{h+e}}^{\mathrm{smearing}};\overline{\alpha_{\mathrm{h+e}}^{\mathrm{smearing}}},\Delta\alpha_{\mathrm{h+e}}^{\mathrm{smearing}}\right) \nonumber\\
  && \times \mathcal{N}\left(\sigma_{\mathrm{s}}^{\mathrm{smearing}};\overline{\sigma_{\mathrm{s}}^{\mathrm{smearing}}},\Delta\sigma_{\mathrm{s}}^{\mathrm{smearing}}\right) \times \mathcal{N}\left(\alpha_{\mathrm{s}}^{\mathrm{smearing}};\overline{\alpha_{\mathrm{s}}^{\mathrm{smearing}}},\Delta\alpha_{\mathrm{s}}^{\mathrm{smearing}}\right), \\
  \mathcal{L}_{\mathrm{syst, fake}} &=& \mathcal{N}\left(r_{\mathrm{ABCD}};1,\Delta r_{\mathrm{ABCD}}\right) \times \mathcal{N}\left(p_0;\overline{p_0},\Delta p_0\right) \times \mathcal{N}\left(p_1;\overline{p_1},\Delta p_1\right),\\
  n^H_{\mathrm{s}}  &=& \mu_{\mathrm{s}} \times \alpha^H_{\mathrm{s}},\\
  n^L_{\mathrm{s}}  &=& \mu_{\mathrm{s}} \times \alpha^L_{\mathrm{s}},\\
  r_{\mathrm{ABCD}} &=& \log \frac{n_{\mathrm{f}}^{H}/n_{\mathrm{Fake CR}}^{H}}{n_{\mathrm{f}}^{L}/n_{\mathrm{Fake CR}}^{L}}.
\end{eqnarray}

The total likelihood $\mathcal{L}_{\mathrm{Total}}$ consists of three terms:
a term for the spectrum shape, $\mathcal{L}_{\mathrm{shape}}$, a term to include
systematic uncertainties except for those related to fake-tracket background,
$\mathcal{L}_{\mathrm{syst}}$, and a term for the fake-tracklet background
uncertainties, $\mathcal{L}_{\mathrm{syst, fake}}$.
The numbers of observed events are represented by $n^R_{\mathrm{obs}}$ and
$n^R_{\mathrm{Fake CR, obs}}$ in the signal region and in the fake control region
respectively, where $R$ is $H$
or $L$, representing the high-\met or the low-\met region.
The expected numbers of events for each component (signal, the sum of hadron and electron,
muon and fake-tracklet background) are represented by $n^R_{\mathrm{s}}$, $n^R_{\mathrm{h+e}}$, $n^R_{\mathrm{\mu}}$ and $n^R_{\mathrm{f}}$ respectively.
The normalisation parameter for the signal component is represented by
$\alpha_{\mathrm{s}}^{R}$.
The expected number of signal events is scaled from $\alpha_{\mathrm{s}}^{R}$
using the relative signal strength $\mu_{\mathrm{s}}$.
The probability density functions of those components are represented by
$\mathcal{F}^R_{\mathrm{s}}$, $\mathcal{F}^R_{\mathrm{h+e}}$,
$\mathcal{F}^R_{\mathrm{\mu}}$ and $\mathcal{F}_{\mathrm{f}}$.
The resolution and slope parameters for the smearing functions are
$\sigma^{\mathrm{smearing}}_{\mathrm{s}}$ and $\alpha^{\mathrm{smearing}}_{\mathrm{s}}$
($\sigma^{\mathrm{smearing}}_{\mathrm{h+e}}$ and
 $\alpha^{\mathrm{smearing}}_{\mathrm{h+e}}$) for signal (sum of hadron and electron).
For the fake-tracklet component, the probability density function is common to the low-\met and high-\met regions.
The parameters of the fake-tracklet $\pt$ spectrum's shape function are represented by $p_0$ and $p_1$.
The fake-tracklet ratio factor between the low-\met and high-\met regions, $r_{\mathrm{ABCD}}$, is
derived from 
$n^H_{\mathrm{f}}$, $n^L_{\mathrm{f}}$ and the expected numbers of events in the fake-tracklet control
regions, $n_{\mathrm{Fake CR}}^{H}$ and $n_{\mathrm{Fake CR}}^{L}$.
The parameters constrained by the fit are:
$\mu_{\mathrm{s}}$,
$n^H_{\mathrm{h+e}}$,
$n^H_{\mathrm{\mu}}$,
$n^H_{\mathrm{f}}$,
$n^H_{\mathrm{Fake CR}}$,
$n^L_{\mathrm{h+e}}$,
$n^L_{\mathrm{\mu}}$,
$n^L_{\mathrm{f}}$,
$n^L_{\mathrm{Fake CR}}$,
$\sigma^{\mathrm{smearing}}_{\mathrm{s}}$,
$\alpha^{\mathrm{smearing}}_{\mathrm{s}}$,
$\sigma^{\mathrm{smearing}}_{\mathrm{h+e}}$,
$\alpha^{\mathrm{smearing}}_{\mathrm{h+e}}$,
$p_0$ and
$p_1$.
Other parameters are fixed in the fit.
The expected value and the uncertainty of a variable $x$ is represented by $\overline{x}$ and $\Delta x$ respectively.
The value of a unit Gaussian-function at $a$ with mean $b$ and standard deviation $c$ is represented by $\mathcal{N}\left(a;b,c\right)$.

%

\printbibliography



\clearpage 
\begin{flushleft}
{\Large The ATLAS Collaboration}

\bigskip

M.~Aaboud$^\textrm{\scriptsize 137d}$,
G.~Aad$^\textrm{\scriptsize 88}$,
B.~Abbott$^\textrm{\scriptsize 115}$,
O.~Abdinov$^\textrm{\scriptsize 12}$$^{,*}$,
B.~Abeloos$^\textrm{\scriptsize 119}$,
S.H.~Abidi$^\textrm{\scriptsize 161}$,
O.S.~AbouZeid$^\textrm{\scriptsize 139}$,
N.L.~Abraham$^\textrm{\scriptsize 151}$,
H.~Abramowicz$^\textrm{\scriptsize 155}$,
H.~Abreu$^\textrm{\scriptsize 154}$
% ...
\bigskip
\\
% ...
$^{12}$ Institute of Physics, Azerbaijan Academy of Sciences, Baku, Azerbaijan\\
% ...
$^{88}$ CPPM, Aix-Marseille Universit{\'e} and CNRS/IN2P3, Marseille, France\\
% ...
$^{115}$ Homer L. Dodge Department of Physics and Astronomy, University of Oklahoma, Norman OK, United States of America\\
% ...
$^{119}$ LAL, Univ. Paris-Sud, CNRS/IN2P3, Universit{\'e} Paris-Saclay, Orsay, France\\
% ...
$^{137}$ $^{(a)}$ Facult{\'e} des Sciences Ain Chock, R{\'e}seau Universitaire de Physique des Hautes Energies - Universit{\'e} Hassan II, Casablanca; $^{(b)}$ Centre National de l'Energie des Sciences Techniques Nucleaires, Rabat; $^{(c)}$ Facult{\'e} des Sciences Semlalia, Universit{\'e} Cadi Ayyad, LPHEA-Marrakech; $^{(d)}$ Facult{\'e} des Sciences, Universit{\'e} Mohamed Premier and LPTPM, Oujda; $^{(e)}$ Facult{\'e} des sciences, Universit{\'e} Mohammed V, Rabat, Morocco\\
% ...
$^{139}$ Santa Cruz Institute for Particle Physics, University of California Santa Cruz, Santa Cruz CA, United States of America\\
% ...
$^{151}$ Department of Physics and Astronomy, University of Sussex, Brighton, United Kingdom\\
% ...
$^{154}$ Department of Physics, Technion: Israel Institute of Technology, Haifa, Israel\\
$^{155}$ Raymond and Beverly Sackler School of Physics and Astronomy, Tel Aviv University, Tel Aviv, Israel\\
% ...
$^{161}$ Department of Physics, University of Toronto, Toronto ON, Canada\\
% ...
$^{a}$ Also at Department of Physics, King's College London, London, United Kingdom\\
$^{b}$ Also at Institute of Physics, Azerbaijan Academy of Sciences, Baku, Azerbaijan\\
$^{c}$ Also at Novosibirsk State University, Novosibirsk, Russia\\
$^{d}$ Also at TRIUMF, Vancouver BC, Canada\\
$^{e}$ Also at Department of Physics {\&} Astronomy, University of Louisville, Louisville, KY, United States of America\\
$^{f}$ Also at Physics Department, An-Najah National University, Nablus, Palestine\\
% ...
$^{h}$ Also at Department of Physics, University of Fribourg, Fribourg, Switzerland\\
% ...
$^{s}$ Also at Department of Financial and Management Engineering, University of the Aegean, Chios, Greece\\
% ...
$^{x}$ Also at Graduate School of Science, Osaka University, Osaka, Japan\\
% ...
$^{*}$ Deceased
\end{flushleft}


\end{document}